\documentstyle[namedreferences,editedvolume,psfig]{crckapb} 

\newcommand       \Angstrom     {\,{\rm \AA}}     
\newcommand       \simlt        {\leq}
\newcommand       \simgt        {\geq}
\newcommand       \gtsim        {\geq}
\newcommand       \ltsim        {\leq}
\newcommand       \magkpc       {\,{\rm mag\, kpc}^{-1}}
\newcommand       \g            {\,{\rm g}}

\newcommand       \nm           {\,{\rm nm}}
\newcommand       \K            {\,{\rm K}}
\newcommand       \km           {\,{\rm km}}
\newcommand       \s            {\,{\rm s}}
\newcommand       \ppm          {\,{\rm ppm}}
\newcommand       \um           {\,{\rm \mu m}}
\newcommand       \micron       {\,{\rm \mu m}}
\newcommand       \umrev        {\,{\rm \mu m^{-1}}}
\newcommand       \cm           {\,{\rm cm}}
\newcommand       \msun         {\,{\rm m_\odot}}
\newcommand       \Rv           {R_V}
\newcommand       \lambdamax    {\lambda_{\rm  max}}
\newcommand       \Pmax         {P_{\rm  max}}
\newcommand       \Ahump        {A_{2175\Angstrom}}
\newcommand       \Phump        {P_{2175\Angstrom}}

\newcommand       \url          {\,{\rm http://}}
  
\begin{opening}
\title{In Dust We Trust: An Overview of Observations and
Theories of Interstellar Dust}
\subtitle{}
\author{Aigen Li}
\institute{Princeton University Observatory, Peyton Hall, 
   Princeton,\\ NJ 08544, USA; {\sf agli@astro.princeton.edu}\\
   and Theoretical Astrophysics Program,
   University of Arizona,\\ Tucson, AZ 85721, USA;
   {\sf agli@lpl.arizona.edu}}
\author{J. Mayo Greenberg}
\institute{The Raymond and Beverly Sackler Laboratory for Astrophysics, 
Sterrewacht Leiden, Postbus 9513, 2300 RA Leiden,\\ The Netherlands}
\end{opening}

\runningtitle{Interstellar Dust}

\begin{document}

\begin{abstract}
The past century of interstellar dust has brought us from 
first ignoring it to finding that it is an important component 
of the interstellar medium and plays an important role 
in the evolution of galaxies, the formation of stars and
planetary systems, and possibly, the origins of life.
Current observational results in our galaxy provide 
a complex physical and chemical evolutionary 
picture of interstellar dust starting with the formation of 
small refractory particles in stellar atmospheres to their 
modification in diffuse and molecular clouds and
ultimately to their contribution to star forming regions.
In this review, a brief history of the studies of interstellar
dust is presented. Our current understanding of the physical
and chemical properties of interstellar dust are summarized, 
based on observational evidences from interstellar extinction, 
absorption, scattering, polarization, emission (luminescence,
infrared vibrational emission, and microwave rotational emission),  
interstellar depletions, and theoretical modelling. 
Some unsolved outstanding problems are listed.
\end{abstract}

\vspace{-5mm}
\section{Introduction}
It has been over 70 years since the existence of solid dust
particles in interstellar space was first convincingly shown
by Trumpler (1930) based on the discovery of color excesses. 
Interstellar dust has now become a subject of extensive study
and one of the subjects in the forefront of astrophysics.

\vspace{2mm}
\hspace{6mm}{\tt ``The role of dust is that of observer and of catalyst.''}

\hspace{68mm}{\sf ------ J. Mayo Greenberg [1963]}
\vspace{2mm}

Historically, interstellar dust was regarded by astronomers
as an annoying interstellar ``fog'' which prevented
an accurate measurement of distances to stars.  
About 40 years ago, one of us (J.M.G.) wrote in the first
volume of {\it Annual Review of Astronomy and Astrophysics}
(Greenberg 1963):\footnote{%
 The contents in the square brackets were added by A. Li,
 for completeness, to summarize the then understanding as
 discussed in Greenberg (1963).
 }
``... Among the various performers of our Galaxy -- the stars,
the gas clouds, the cosmic rays -- the grains seem to be the 
least dramatic. Their role is generally that of observer and 
of catalyst of events rather than prime mover. Why then are we
so interested in these small particles whose total mass, by the
most generous estimate, is only of the order of 1 per cent of
that of the gas clouds? ... [It is because of] the three important
activities of the grains: (a) the negative one of extinction
[blocking the light from distant stars];
(b) the positive one of tracer of physical conditions
[e.g. the Galactic magnetic fields and the gas temperature];
and (c) physical interactions with other components of 
the interstellar medium [e.g. the formation of molecules and stars].''

\vspace{2mm}
\hspace{2mm}{\tt ``We now recognize, dust plays a role not only 
as a tracer}\\
\vspace{-1mm} 
\hspace{8mm}{\tt of what goes on in space, a corrector for making 
different}\\
\vspace{0mm} 
\hspace{8mm}{\tt modifications in our idea of the morphology of
galaxies,}\\
\vspace{0mm} 
\hspace{8mm}{\tt but also actively contributing to the chemical
evolution}\\
\vspace{0mm} 
\hspace{8mm}{\tt of molecular clouds.''}

\hspace{68mm}{\sf ------ J. Mayo Greenberg [1996]}
\vspace{2mm}

It is seen now that the role of interstellar dust was significantly
underestimated 40 years ago. The advances of infrared (IR) astronomy, 
ultraviolet (UV) astronomy, laboratory astrophysics, and theoretical
modelling over the past 40 years have had a tremendous impact 
on our understanding of the physical and chemical nature, origin 
and evolution of interstellar grains and their significance in 
the evolution of galaxies, the formation of stars and stellar systems 
(planets, asteroids, and comets), and the synthesis of complex organic
molecules which possibly leads to the origins of life. 

\vspace{2mm}
\hspace{-10mm}{\tt ``Dust is both a subject and an agent of 
the Galactic evolution.''}

\hspace{52mm}{\sf ------ J. Dorschner \& Th. Henning [1995]}
\vspace{2mm}

Instead of being a passive ``observer'', interstellar dust 
plays a vital role in the evolution of galaxies.      
Besides providing $\sim 30\%$ of the total Galactic luminosity 
via their IR emission, dust grains actively participate in 
the cycle of matter (gas and dust) from the interstellar 
medium (ISM) to stars and back from stars to the ISM: 
(1) solid grains condense in the cool atmospheres of 
evolved stars, Wolf-Rayet stars, planetary nebulae, and
novae and supernovae ejecta, and are then ejected into 
the diffuse ISM;
(2) in the diffuse ISM, interacting with hot (shocked) gas, 
stellar UV radiation, and cosmic rays, grains undergo 
destruction (sputtering by impacting gas atoms, vaporization
and shattering by grain-grain collisions; see Tielens 1999 
for a review);
(3) in molecular clouds (formed through shock compression
of diffuse gas, agglomeration of small clouds and 
condensation instabilities), grains are subject to growth 
through accretion of an ice mantle and coagulation; 
(4) cycling between the diffuse and molecular clouds, 
dust grains either form a carbonaceous organic refractory 
mantle as a result of UV processing of the ice mantle
accreted on the silicate core in molecular clouds 
(Greenberg et al.\ 1972; Greenberg \& Li 1999a), 
or perhaps grains re-condense in dense regions followed by rapid 
exchange of matter between diffuse gas and dense gas (Draine 1990); 
(5) collapse of dense molecular clouds leads to the birth
of new stars. At the late stages of stellar evolution, 
gas and newly formed dust will eventually return to the ISM 
either through stellar winds or supernova explosions.
It is clear that the life cycle of dust is associated
with that of stars; stars are both a sink and a major
source for the Galactic dust.

In addition to the fact that stars form out of interstellar 
dust and gas clouds, dust plays an important role in the 
process of star formation in molecular clouds: 
(1) IR emission from dust removes the gravitational energy 
of collapsing clouds, allowing star formation to take place;
(2) dust grains provide shielding of molecular regions 
from starlight and thereby reduce the ionization levels 
and speed up the formation of protostellar cores 
(Ciolek 1995);\footnote{%
 Molecular clouds are partially supported by magnetic fields;
 star formation occurs if ambipolar diffusion deprives 
 cloud cores of magnetic support.
 Charged grains can couple to the magnetic field and 
 increase the collisional drag on the neutrals, 
 and thereby slow the rate of ambipolar diffusion 
 within a cloud and increase the time needed to 
 form a protostellar core (Ciolek 1995).
 }
(3) IR emission from dust provides an effective 
probe for the star-formation processes (Shu, Adams, \& Lizano 1989).

\vspace{2mm}
\hspace{-6mm}{\tt ``The importance of grains in various aspects of
astrochemistry\\ is evident: they shield molecular regions from 
dissociating in-\\-terstellar radiation, catalyze formation of
molecules, and re-\\-move molecules from the gas phase.''}

\hspace{5mm}{\sf ------ E.F. van Dishoeck, G.A. Blake, B.T. Draine,
\& J.I. Lunine [1993]}
\vspace{2mm}

The interstellar chemistry problem concerns the chemical 
reactions between dust and atoms and molecules in space;
dust plays an active role in those reactions 
(see van Dishoeck et al.\ 1993, van Dishoeck 1999 for reviews): 
(1) grain surfaces provide the site for the formation
of molecular hydrogen (see Pirronello 2002 for a review), 
and probably other simple molecules through grain surface 
reactions and complex organic molecules through UV photoprocessing 
(e.g. see Greenberg et al.\ 2000; Allamandola 2002);
(2) dust reduces the stellar UV radiation and protects
molecules from photodissociation;
(3) dust provides the major heating source for interstellar
gas -- photoelectrons ejected from grains;
(4) dust grains are also involved in ion-molecule chemistry
by affecting the electron/ion densities within the cloud.

Dust is one of the basic ingredients in comets. 
There is growing evidence from cometary observations 
that comets are a storage place for products of the chemical 
evolution which takes place in interstellar space. 
The complex chemistry and molecular evolution leading to 
what is now seen in comets may be the necessary precursor 
to life on the Earth and there is reason to believe from 
the evidence available that the oceans on the Earth were 
made of comets bringing interstellar ice to our young planet 
(Ehrenfreund \& Charnley 2000).

In this review, we start in \S2 with a historically oriented
discussion of the discovery of interstellar extinction and dust, 
the development of early dust models as well as current modern models.
Following up in \S3 we present the present state-of-the-art 
understanding of interstellar dust observations and theories.
In \S4 we present a personal perspective of future dust studies.

\section{History of Dust Studies}

\subsection{Interstellar Dust: Early Observational Evidences}

\vspace{2mm}
\hspace{40mm}{\tt ``Surely, there is a hole in the heavens!''}

\hspace{65mm}{\sf ------ Sir William Herschel [1785]}
\vspace{2mm}

The Milky Way looks patchy with stars unevenly distributed:
looking at the sky in the direction of Sagittarius 
it is clear that there are tremendously dark lanes especially in 
the region toward the Galactic Center. 
The subject of these dark patches and what makes them dark also 
have a very patchy history. The existence of dark regions in the
Milky Way was first pointed out by Sir  William Herschel
in the late 18th century. At that time, these dark lanes -- 
the dust clouds which obscure the light from the background 
stars, were considered as ``holes in the heavens'' (Herschel 1785).

\vspace{2mm}
\hspace{46mm}{\tt ``They are really Obscuring bodies!''}

\hspace{77mm}{\sf ------ Agnes Clerke [1903]}
\vspace{2mm}

In August 1889, Edward Barnard started to take pictures 
and reported vast and wonderful cloud forms with their remarkable 
structure, lanes, holes and black gaps.
At the beginning of the 20th century, astronomers started to 
realize that they ``were really obscuring bodies'' rather than 
holes devoid of stars (Barnard 1919). 
Agnes Clerke (1903) stated in an astrophysics text that 
``... The fact is a general one, that in all the forest of 
the universe there are glades and clearings. How they
come to be thus diversified we cannot pretend to say; 
but we can see that the peculiarity is structural ---
that it is an outcome of the fundamental laws governing 
the distribution of cosmic matter. Hence the futility of 
trying to explain its origin, as a consequence, for instance, 
of the stoppage of light by the interposition of obscure bodies, 
or aggregations of bodies, invisibly thronging space.''

Heber D. Curtis and Harlow Shapley\footnote{%
 Henry Norris Russell (1922), Shapley's advisor at Princeton,
 believed that the existence of dark clouds accounted for
 the obscuration and argued that this obscuring matter had to
 be the form of fine dust. But Shapley did not follow
 his advice.
 }
held a famous debate in 1920 (Shapley \& Curtis 1921);
among the points of contention was whether what is seen as
the dark lanes in the Milky Way is caused by obscuring material. 
Curtis said the dark lanes observed in our Galaxy were obscuring 
material, while Shapley said he found no evidence of obscuring 
material in his observations of globular clusters. 
Later observers became aware that Shapley's argument was
irrelevant because the globular clusters are out of the plane 
of the Galaxy. The obscuring dust was confined to the so called 
``plane of avoidance'' which is the Galactic plane.

\vspace{2mm}
\hspace{78mm}{\tt ``Stars are dimmed!''}

\hspace{73mm}{\sf ------ Wilhelm Struve [1867]}
\vspace{2mm}

The presence of interstellar extinction was pointed out as early as 
in 1847 by F.G. Wilhelm Struve. He found that the number of stars 
per unit volume seems to diminish in all directions receding 
from the Sun. This could be explained either if the Sun was 
at the center of a true stellar condensation, or if the effect 
was only an apparent one due to absorption (which may have been 
understood to include light scattering). He argued that there 
could be an visual extinction of about $1\magkpc$ in interstellar 
space.

Jacobus C. Kapteyn (1904) had found a roughly spherical 
distribution of stars around the Sun. 
He assumed a constant stellar density and then 
used the observed density to arrive at a value for the extinction 
(absorption) of light $\approx 1.6\magkpc$ (Kapteyn 1909), 
which differs little from current values 
($\approx 1.8\magkpc$ assuming a hydrogen
density of $n_{\rm H} = 1\cm^{-3}$).\footnote{%
 But Kapteyn did not take this seriously; 
 for example, he assumed no extinction in his grand
 1922 paper on the motion of stars in the Galaxy (Kapteyn 1922).
 }

In  1929 Schal\'{e}n examined the question of
stellar densities as a function of distance. 
He did a very detailed study of B and A stars, 
including those in Cygnus, Cepheus, Cassiopeia, 
and Auriga. He obtained rather different values
of the absorption coefficient, particularly in 
Cygnus and Auriga where there are large dark patches. 
So obviously the absorption is more in some regions 
and less in others.

\vspace{2mm}
\hspace{2mm}{\tt ``Cosmic dust particles produce the selective absorption.''}

\hspace{66mm}{\sf ------ Robert J. Trumpler [1930]}
\vspace{2mm}

It was not until the work of Robert J. Trumpler in 1930 that 
the first evidence for interstellar reddening was found. 
Trumpler (1930) based this on his study of open clusters 
in which he compared the luminosities and distances of 
open clusters with the distances obtained by assuming that 
all their diameters were the same. By observing the luminosities 
and knowing the spectral distribution of stars he was able 
to find both absorption ($\approx 0.7\magkpc$) 
and selective absorption or color excess 
(between photographic and visual; $\approx 0.3\magkpc$)
with increasing distance,
and produce a reddening curve.\footnote{%
  Trumpler's observations indicated reddening even 
  where he saw no clouds.
  Dufay (1957) questioned whether interstellar space 
  outside dark clouds and nebulae should be considered 
  perfectly transparent.} 
It was this work which led to the general establishment of 
the existence of interstellar dust. 

The wavelength dependence of extinction in the optical
was measured for the first time by Rudnick (1936) using
the still-widely-used ``pair-match'' method.
Further observations carried out by Hall (1937)
and Stebbins, Huffer \& Whitford (1939) pointed
to an $\lambda ^{-1}$ reddening ``law''
(at that time limited to $1-3\umrev$):
the reddening curve showed a rise inversely 
proportional to the wavelength $\lambda$.

In 1934 Paul W. Merrill reported the discovery of
the 5780$\Angstrom$, 5797$\Angstrom$, 6284$\Angstrom$, 
6614$\Angstrom$ ``unidentified interstellar lines''. 
These widened absorption lines, now known as 
``Diffuse Interstellar Bands'', still remain 
unidentified (see Krelowski 2002 for a review).

The existence of diffuse interstellar radiation,
originally detected by van Rhijn (1921), was verified
and attributed to small dust grains by Henyey \& Greenstein (1941). 
Henyey \& Greenstein (1941) found that interstellar
particles are strongly forward scattering and have a high albedo.

\vspace{2mm}
\hspace{65mm}{\tt ``Starlight is polarized!''}

\hspace{44mm}{\sf ------ John S. Hall [1949]; W.A. Hiltner [1949]}
\vspace{2mm}

At the end of 1940s, two investigators (Hall 1949; Hiltner 1949), 
inspired by a prediction of Chandrasekhar on 
intrinsic stellar polarization,
independently discovered instead the general interstellar 
linear polarization. Magnetic fields were
believed to confine cosmic rays and to play a role 
in the spiral structure of the Galaxy. The implication of
the linear polarization was that the extinction was 
caused by non-spherical particles aligned by magnetic fields
(Davis \& Greenstein 1951). 
The wavelength dependent polarization curve was later shown 
to be well represented by the Serkowski law, an empirical 
formula (Serkowski 1973). In addition to the extinction curve 
which was later also extended to a wide wavelength range, 
the polarization law as well as the polarization 
to extinction ratio provide further insight into 
the physical and chemical nature of interstellar dust.

The circular polarization produced by interstellar
birefringence (Martin 1972) was originally predicted 
by van de Hulst (1957). It was first detected along
the lines of sight to the Crab Nebula by Martin, Illing, 
\& Angel (1972) and to six early-type stars by 
Kemp \& Wolstencroft (1972).  

\subsection{Interstellar Matter: Theoretical Evidences}

\vspace{2mm}
\hspace{32mm}{\tt ``There must be more matter than stars......''}

\hspace{78mm}{\sf ------ Jan H. Oort [1932]}
\vspace{2mm}

In the 1930's Jan H. Oort took another approach to the problem 
by looking at the statistics of the motions of K giants 
perpendicular to the plane of the Galaxy, that is, at bulge objects. 
He used these to estimate the mass of material in the plane. 
He found that there had to be more material there than could
be seen in stars. Oort (1932) estimated that the mass of 
the non-stellar material 
(dust and gas) is about $12\times10^9\msun$. 
If this mass is distributed uniformly, 
the density of this non-stellar material is 
$\rho_{\rm ism} \approx 6 \times 10^{-24}\g\cm^{-3}$ --
this is the mass required to explain the observed motions.

The question then becomes what kind of material distributed 
with this density with what mass absorption coefficient could 
give rise to an extinction of about $1\magkpc$, as observed. 
So what is required is that the scattering/extinction 
cross section of the material blocking the light per unit length 
is on the order of $1\magkpc$.

\subsection{Interstellar Dust: Early Modelling Efforts}

\vspace{2mm}
\hspace{80mm}{\tt 1930s: Metallic grains}

\hspace{66mm}{\sf ------ C. Schal\'{e}n; J.L. Greenstein}
\vspace{2mm}

But what caused the interstellar reddening? 
Since hyperbolic meteors were thought to exist, 
first attempts were made to tie the interstellar dust 
to the meteors. Small metallic particles were among 
the materials initially proposed to be responsible for 
the interstellar reddening, based on an analogy with small
meteors or micrometeorites supposedly fragmented into 
finer dust (Schal\'{e}n 1936; Greenstein 1938).\footnote{%
 Greenstein (1938) concluded that a dust size distribution
 of $dn/da \sim a^{-3.6}$ ``seems to provide the best
 agreement of theory and observation (interstellar extinction).''
 Interesting enough, this power law distribution was very
 close to the $dn/da \sim a^{-3.5}$ distribution derived
 about 40 years later for the silicate/graphite dust model
 (Mathis, Rumpl, \& Nordsieck 1977; Draine \& Lee 1984).   
 }
Reasonably good fits to the $\lambda^{-1}$ extinction law 
were obtained in terms of small metallic grains with sizes 
of the order of $0.01\um$.\footnote{%
  Another possible influencing factor of proposing the 
  metallic dust model might have been the fact that 
  it was easier to compute the scattering by metallic 
  particles using the Mie theory because to get 
  a $\lambda ^{-1}$ law required smaller particles than 
  if they were dielectric and computations for large particles 
  were too tedious (van de Hulst 1986).}
It became evident later that meteors or micrometeorites 
are not of interstellar origin.

In 1948 Whitford published measurements of star colors 
versus spectral types over a wavelength range from about 
3500{\AA} (UV) to the near IR. The relation 
was not the expected straight line, but showed curvature 
at the near UV and IR regions. 
Things were beginning to make some physical sense 
from the point of view of small particle scattering.

\vspace{2mm}
\hspace{72mm}{\tt 1940s: Dirty ice grains}

\hspace{60mm}{\sf ------ J.H. Oort \& H.C. van de Hulst}
\vspace{2mm}

Based on the correlation between gas concentration and 
extinction, Lindblad (1935) argued that it seemed reasonable 
to grow particles in space since, as hypothesized by 
Sir Arthur Eddington, gaseous atoms and ions which hit 
a solid particle in space would freeze down upon it.  
Lindblad (1935) further put forward the hypothesis that 
interstellar dust could have formed by condensation (or more
properly, accretion) of interstellar gas. 

In the 1940's van de Hulst (1949) broke with tradition and 
published the results of making particles out of atoms that 
were known to exist in space: H, O, C, and N. 
He assumed these atoms combined on the surface to 
form frozen saturated  molecules. 
The gas condensation scenario was further investigated 
by Oort \& van de Hulst (1946) and led to what later 
became known as the ``dirty ice'' model.\footnote{%
 A. Li noted: the term ``dirty ice'' was invented by J.M. Greenberg
 as recalled by H.C. van de Hulst (1997). 
 }

The dirty ice model of dust by van de Hulst was a logical 
followup of the then existing information about 
the interstellar medium and contained the major idea 
of surface chemistry leading to the ices H$_2$O, CH$_4$, NH$_3$. 
But it was not until the advent of IR astronomical techniques 
made it possible to observe silicate particles emitting at their
characteristic 10$\um$ wavelength in the atmospheres of cool stars 
that we had the cores on which the matter could form. 
Interestingly, their presence was predicted on theoretical 
grounds by Kamijo (1963). As van de Hulst said, he chose to
ignore the nucleation problem and just go ahead 
(where no one had gone before) with the assumption
that ``something'' would provide the seeds for the mantle 
to grow on. By 1945 we had many of the theoretical basics 
to understand the sources of interstellar dust ``ices'' 
but it was not until about 1970 that the silicates were 
established. Without having a realistic dust model, 
van de Hulst developed the scattering tools to 
provide a good idea of dust properties.

\vspace{2mm}
\hspace{74mm}{\tt 1960s: Graphite grains}

\hspace{55mm}{\sf ------ F. Hoyle \& N.C. Wickramasinghe}
\vspace{2mm}

A ``challenge'' to the dirty ice model came out 
just after the discovery of interstellar polarization 
(Hall 1949; Hiltner 1949) since it seemed that 
the dirty ice model could not explain the rather 
high degree of polarization relative to extinction
(see van de Hulst 1957; this was later shown not to 
be true by Greenberg et al.\ 1963a, b). 

This led to the re-consideration of metallic grains
and the consideration of graphite condensed 
in the atmospheres of carbon stars as a dust component 
(Cayrel \& Schatzman 1954; Hoyle \& Wickramasinghe 1962)
because of their enormous potential for polarizing stellar 
radiation (as a result of its anisotropic optical properties
of graphite).
The graphite proposal seemed to be further supported by 
the detection of the 2175$\Angstrom$ hump extinction 
(Stecher \& Donn 1965), although we know nowadays that 
the graphite model is not fully successful in 
explaining the 2175$\Angstrom$ hump and the ultimate 
identification is still not made (see \S3.1.2). 

Kamijo (1963) first proposed that SiO$_2$, condensed 
in the atmospheres of cool stars and blown out into 
the interstellar space, could provide condensation cores 
for the formation of ``dirty ices''. It was later shown 
by Gilman (1969) that grains around oxygen-rich cool giants 
are mainly silicates such as Al$_2$SiO$_3$ and Mg$_2$SiO$_4$. 
Silicates were first detected in emission in M stars 
(Woolf \& Ney 1969; Knacke et al.\ 1969a), in the Trapezium 
region of the Orion Nebula (Stein \& Gillett 1969), and in comet 
Bennett 1969i (Maas, Ney, \& Woolf 1970); in absorption toward the 
Galactic Center (Hackwell, Gehrz, \& Woolf 1970), and toward the
Becklin-Neugebauer object and Kleinmann-Low Nebula (Gillett 
\& Forrest 1973). Silicates are now known to be ubiquitous, seen in 
interstellar clouds, circumstellar disks around young stellar objects
(YSOs), main-sequence stars and evolved stars, in HII regions, and in 
interplanetary and cometary dust (see Li \& Draine 2001a for a review).

\subsection{Contemporary Interstellar Dust Models}

\vspace{0mm}
\hspace{70mm}{\tt Since 1970s: Modern models}

\hspace{6mm}{\sf ------ J.M. Greenberg; J.S. Mathis; B.T. Draine;
and their co-workers}
\vspace{1mm}

The first attempt to find the 3.1$\um$ feature of H$_2$O was 
unsuccessful (Danielson, Woolf, \& Gaustad 1965; Knacke, Cudaback, 
Gaustad 1969b). This was, at first, a total surprise to those who 
had accepted the dirty ice model. 
However, this gave the incentive to perform the early experiments 
on the UV photoprocessing of low temperature mixtures of 
volatile molecules simulating the ``original'' dirty ice grains 
(Greenberg et al.\ 1972; Greenberg 1973) to understand how and why 
the predicted H$_2$O was not clearly present. 
From such experiments was predicted a new component 
of interstellar dust in the form of complex organic 
molecules, as mantles on the silicates.
This idea was further developed in the framework of the cyclic 
evolutionary silicate core-organic refractory mantle dust model 
(Greenberg 1982a; Greenberg \& Li 1999a). 
Similar core-mantle models have also been proposed 
by others (D\'{e}sert, Boulanger, \& Puget 1990;
Duley, Jones, \& Williams 1989; Jones, Duley, \& Williams 1990).

According to the cyclic evolutionary model, ices evolve 
chemically and physically in interstellar space, 
so do the organics. 
Where and how the interstellar dust is formed appears to 
involve a complex evolutionary picture. The rates of 
production of refractory components such as silicates 
in stars do not seem to be able to provide more than 
about 10\% of what is observed in space because they 
are competing with destruction which is about 10 times 
faster by, generally, supernova shocks 
(Draine \& Salpeter 1979a,b; Jones et al.\ 1994). 
At present the only way to account for the observed 
extinction amount is to resupply the dust by 
processes which occur in the interstellar medium
itself. The organic mantles on the silicate particles must 
be created at a rate sufficient to balance their
destruction. Furthermore, they provide a shield against 
destruction of the silicates. Without them the
silicates would indeed be underabundant 
unless most of the grain mass was condensed 
in the ISM, as suggested by Draine (1990).

What is currently known about the organic dust component 
is based very largely on results of laboratory experiments 
which attempt to simulate interstellar processes. 
The organic refractories which are derived from 
the photoprocessing of ices contain a mixture of aliphatic 
and aromatic carbonaceous molecules (Greenberg et al.\ 2000). 
The laboratory analog suggests the presence of abundant 
prebiotic organic molecules in interstellar dust 
(Briggs et al.\ 1992). 

The silicate core-organic mantle model is recently revisited
by Li \& Greenberg (1997) in terms of a trimodal size 
distribution consisting of (1) large core-mantle grains
which account for the interstellar polarization,
the visual/near-IR extinction, and the far-IR emission;
(2) small carbonaceous grains of graphitic nature to
produce the 2175$\Angstrom$ extinction hump;
(3) polycyclic aromatic hydrocarbons (PAHs) to
account for the far-UV extinction as well as the
observed near- and mid-IR emission features at
3.3, 6.2, 7.7, 8.6, and 11.3$\um$. This model is 
able to reproduce both the interstellar extinction 
and linear and circular polarization.

An alternative model was proposed by Mathis, Rumpl, 
\& Nordsieck (1977) and thoroughly extended by 
Draine \& Lee (1984). It consists of two separate dust 
components -- bare silicate and graphite particles.
Modifications to this model was later made by 
Sorrell (1990), Siebenmorgen \& Kr\"{u}gel (1992), 
and Rowan-Robinson (1992) by adding new dust components
(amorphous carbon, PAHs) and changing dust sizes. 

Very recently, Draine and his co-workers 
(Li \& Draine 2001b, 2002a; Weingartner \& Draine 2001a) 
have extended the silicate/graphite grain model 
to explicitly include a PAH component as 
the small-size end of the carbonaceous grain population. 
The silicate/graphite-PAHs model provides an excellent 
quantitative agreement with the observations of IR emission
as well as extinction from the diffuse ISM of the Milky Way Galaxy 
and the Small Magellanic Cloud. 

Mathis \& Whiffen (1989) have proposed that interstellar
grains are composite collections of small silicates, 
vacuum ($\approx 80\%$ in volume), 
and carbon of various kinds (amorphous carbon, hydrogenated 
amorphous carbon, organic refractories). 
However, the composite grains may be too cold and produce 
too flat a far-IR emissivity to explain the observational 
data (Draine 1994).\footnote{%
  Let $C_{\rm abs}(\lambda) \propto \lambda^{-\beta}$ be 
  the far-IR absorption cross section; $T_{\rm d}$ be the 
  characteristic dust temperature; $j_{\lambda} \propto 
  C_{\rm abs}(\lambda) \times 4\pi B_{\lambda}(T_{\rm d})
  \propto \lambda^{-(4+\beta)}$ be the dust far-IR emissivity
  (where $B_{\lambda}[T_{\rm d}]$ is the Planck function at 
   wavelength $\lambda$ and temperature $T_{\rm d}$). 
  While the observed emission spectrum between 100$\um$
  and 3000$\um$ (Wright et al.\ 1991; Reach et al.\ 1995)
  is well represented by dust with 
  $\beta=1.7$, $T_{\rm d}\approx 19.5\K$ 
  or $\beta=2.0$, $T_{\rm d}\approx 18.5\K$ (Draine 1999),
  fluffy composite grains have $\beta \approx 1.60$ 
  (Mathis \& Whiffen 1989). A $\beta=2.0$ emissivity law 
  is naturally expected for solid compact dust: 
  the far-IR absorption formula for spherical 
  submicron-sized grains is
  $C_{\rm abs}(\lambda)/V = 18\pi/\lambda\times 
  \left\{\epsilon_{\rm im}/\left[\left(\epsilon_{\rm re}+2\right)^2+
  \epsilon_{\rm im}^2\right]\right\}$ where $V$ is the grain volume;
  $\epsilon_{\rm re}$ and $\epsilon_{\rm im}$ are respectively 
  the real and imaginary part of the dielectric function. 
  For dielectrics $\epsilon_{\rm im} \propto \lambda^{-1}$
  and $\epsilon_{\rm re} \propto {\rm const}$ 
  ($\epsilon_{\rm im} \ll \epsilon_{\rm re}$)
  while for metals $\epsilon_{\rm im} \propto \lambda$,
  $\epsilon_{\rm re} \propto {\rm const}$ 
  ($\epsilon_{\rm im} \gg \epsilon_{\rm re}$)
  so that one gets the same asymptotic relation 
  for both dielectrics and metals
  $C_{\rm abs}(\lambda) \propto \lambda^{-2}$.
  }
This is also true for the fractal grain 
model (Wright 1987).  

In view of the recent thoughts that the reference abundance of
the ISM (the abundances of heavy elements in both solid and
gas phases) is subsolar (Snow \& Witt 1995, 1996),
Mathis (1996, 1998) updated the composite grain model
envisioned as consisting of three components: 
(1) small silicate grains to produce the far-UV 
($\lambda^{-1}>6\umrev$) extinction rise; 
(2) small graphitic grains to produce the 2175$\Angstrom$
extinction hump; (3) composite aggregates of small silicates,
carbon, and vacuum ($\approx 45\%$ in volume) to account for
the visual/near-IR extinction.
The new composite model is able to reproduce the interstellar
extinction curve and the 10$\um$ silicate absorption feature.
But it produces too much far-IR emission in comparison with
the observational data (Dwek 1997).  

\subsection{Scattering of Light by Small Particles: 
An Essential Tool for Dust Studies}
Our knowledge about dust grains is mainly inferred from
their interaction with starlight: a grain in the line 
of sight between a distant star and the observer reduces 
the starlight by a combination of scattering and absorption;
the absorbed energy is then re-radiated in the IR. 
For a non-spherical grain, the light of distant stars 
is polarized as a result of differential extinction for 
different alignments of the electric vector of the radiation.
Therefore, to model the observed interstellar extinction, 
scattering, absorption, polarization and IR emission properties,
knowledge of the optical properties (extinction, absorption 
and scattering cross sections) of interstellar dust is essential. 
This requires knowledge of the optical constants of
the interstellar dust materials [i.e., the complex index of refraction 
$m(\lambda)=m^{\prime}(\lambda)-i\, m^{\prime\prime}(\lambda)$],
and the dust sizes and shapes. Phrasing this differently,
to infer the size, morphology, and chemical 
composition of interstellar dust, an important aspect 
of the modelling of interstellar grains involves 
the computation of extinction, absorption and scattering 
cross sections of particles comprised of candidate materials
and in comparison with astronomical data.

During the recent years, dramatic progress has been made 
in measuring the complex refractive indices of cosmic dust
analogues, e.g. by the Jena Group (Henning \& Mutschke 2000)
and the Naples Group (Colangeli et al.\ 1999).
Interstellar particles would in general be expected to 
have non-spherical, irregular shapes. 
However, our ability to compute scattering and absorption 
cross sections for nonspherical particles is extremely limited. 
So far, exact solutions of scattering problems exist only
for bare or layered spherical grains 
(``Mie theory''; Mie 1908; Debye 1909),
infinite cylinders (Lind \& Greenberg 1966),
and spheroids (Asano \& Yamamoto 1975; Asano \& Sato 1980;
Voshchinnikov \& Farafonov 1993). 
For grains with sizes much smaller than the wavelength
of the incident radiation, the dipole approximation can 
be used to evaluate cross sections for bare or coated
spheroidal grains (van de Hulst 1957; Gilra 1972; 
Draine \& Lee 1984; Bohren \& Huffman 1983). 
The ``T-matrix'' (transition matrix) method, originally
developed by Barber \& Yeh (1975) and recently substantially
extended by Mishchenko, Travis, \& Mackowski (1996),
is able to treat axisymmetric (spheroidal or finite
cylindrical) grains with sizes comparable to the wavelength.
The discrete dipole approximation (DDA), originally developed
by Purcell \& Pennypacker (1973) and recently greatly
improved by Draine (1988), is a powerful technique for 
irregular heterogeneous grains with sizes as large as
several times the wavelength. The VIEF (volume integration of
electric fields) method developed by Hage \& Greenberg (1990), 
based on an integral representation of Maxwell's equations, is
physically similar to the DDA method.
The microwave analog methods originally developed by
Greenberg, Pedersen \& Pedersen (1960) provide 
an effective experimental approach to complex particles.
This method is still proving to be powerful 
(Gustafson 1999; Gustafson et al.\ 1999)
as the needs still outstrip the capacities of computers.

Although interstellar particles are obviously 
non-spherical as evidenced by the observed polarization 
of starlight, the assumption of spherical shapes 
(together with the Bruggeman or the Maxwell-Garnett 
effective medium theories for inhomogeneous grains; 
Bohren \& Huffman 1983) is usually sufficient in modelling 
the interstellar absorption, scattering and IR emission. 
For IR polarization modelling, the dipole approximation 
for spheroidal grains is proven to be successful in many cases.  
The DDA method is highly recommended for studies of 
inhomogeneous (e.g. coated) grains and irregular grains 
such as cometary, interplanetary, 
and protoplanetary dust particles.

\subsection{Comets}
The origin of comets is closely linked to the solar system 
and they played an important role in cosmogony. 
Cometary nuclei have been created far away from the
early Sun\footnote{%
  The current view is that Jupiter family comets (with small 
  inclinations and orbital periods $P<20$ yrs) 
  formed in the trans-Neptune region now known as 
  the ``Kuiper Belt'';
  Halley-type comets (with relatively longer periods $20<P<200$ yrs 
  and larger inclinations) as well as long-period comets (with 
  all possible inclinations and orbital periods $200 <P < 10^7$ yrs)
  formed somewhere beyond the orbits of Jupiter and Saturn. 
  See Li \& Greenberg (1998a) and references therein.
  }
and have been mostly kept intact since their formation. 
Although it is generally accepted that comets 
contain the most pristine material in the solar system, 
it is still a matter of considerable debate whether they 
are made of unmodified protosolar nebula interstellar dust 
or this material has been (completely or partially) evaporated 
before becoming a part of comets 
(see Mumma, Stern, \& Weissman 1993; Crovisier 1999; 
Irvine \& Bergin 2000 for reviews).

The internal structure and chemical composition of 
cometary nuclei have been a topic receiving much attention.
Before 1950, the prevailing view was that the comet nucleus
is composed of a coherent swarm of meteoroidal-type particles 
which are independent of each other 
(termed the ``flying sand-bank'' model).
It was shown by Russell, Dugan, \& Stewart (1926) that 
the ``swarm'' concept of the cometary nucleus was inadequate
since the solar heating would vaporize icy bodies up 
to 30$\cm$ in diameter.

Since 1950 a number of models for the comet nucleus have 
been proposed. For the most part, the icy-conglomerate model 
(also known as the ``dirty snowball'' model) of Whipple (1950) 
has been the standard which others have followed. 
According to Whipple (1950), the comet nucleus is a solid body 
consisting of a conglomerate of refractory dust grains
and frozen gases (mostly of H$_2$O ice). 
More recently, the computer simulation of dust aggregates 
formed by random accumulation (Daniels \& Hughes 1981) led 
Donn, Daniels, \& Hughes (1985) to postulate the fractal model 
in which the comet nucleus is considered as a heterogeneous 
aggregate of ice and dust grains with substantial voids. 
Weissman (1986) proposed a primordial rubble-pile model 
as a modification of the basic icy-conglomerate model in 
which the cometary nucleus is envisaged as a loosely bound 
agglomeration of smaller fragments, weakly bonded by local 
melting at contact interfaces. 
Gombosi \& Houpis (1986) suggested an icy-glue model. 
According to them, the comet nucleus is composed of rather 
large porous refractory boulders (tens of centimeters to 
hundreds of meters) ``cemented'' together with the 
icy-conglomerate type ice-dust grain mix (``Whipple glue'').

Alternatively, Greenberg (1982b) proposed the interstellar 
dust model of comets in which the basic idea is that comets 
have formed directly through coagulation of interstellar dust 
(Greenberg 1982b; Greenberg 1998; Greenberg \& Li 1999b). 
The morphological structure of comet nuclei is thus modelled 
as an aggregate of presolar interstellar dust grains 
whose mean size is of the order of one tenth micron. 
The representative individual presolar grain consists of 
a core of silicates mantled first by an organic refractory material 
and then by a mixture of water dominated ices in which are embedded 
thousands of very small carbonaceous particles/large molecules 
(Greenberg 1998; Greenberg \& Li 1999b). 
Greenberg and his co-workers have further shown how the IR 
emission for several distinctly different types of comets 
bear a general resemblance to each other by reproducing 
the IR emission of various comets 
(Halley -- a periodic comet [Greenberg \& Hage 1990;
Greenberg et al.\ 1996]; Borrelly -- a Jupiter family 
short period comet [Li \& Greenberg 1998a]; 
Hale-Bopp -- a long period comet [Li \& Greenberg 1998b]; 
and extra-solar comets in the $\beta$ Pictoris disk 
[Li \& Greenberg 1998c]) within the framework that 
all comets are made of aggregated interstellar dust.

\section{The Current State of the Art}
The last 40 years have seen a revolution in the study of 
interstellar dust. This has been a four-fold process. 
First of all, the observational access to the UV and 
the IR brought into focus the fact that there had to 
be a very wide range of particle sizes and types to 
account for the blocking of the starlight. 
Secondly, the IR provided a probe of some of 
the chemical constituents of the dust.
Thirdly, laboratory techniques were applied to 
the properties and evolution of possible grain materials. 
Fourthly, advances in numerical techniques and 
the speed and memory of computers have greatly 
enhanced our modelling capabilities.

Thanks to the successful performances of 
IUE ({\it International Ultraviolet Explorer}), 
IRAS ({\it Infrared Astronomical Satellite}), 
COBE ({\it Cosmic Background Explorer}), 
HST ({\it Hubble Space Telescope}), 
ISO ({\it Infrared Space Observatory}) 
as well as various ground-based UV, optical and IR instruments,
we have witnessed an explosive accumulation of new 
observational information on interstellar extinction, polarization, 
scattering, IR continuum emission and spectral features, as well as 
elemental depletion. Rapid progress on laboratory 
experiments has also been made. 
Below we present an overview of our current
knowledge of the dust which brings many related astrophysical 
problems to the fore. 

\subsection{Extinction (Scattering, Absorption) and Polarization}
\subsubsection{Interstellar Extinction and Polarization Curves}

\vspace{2mm}
\hspace{25mm}{\tt ``The extinction curve is a sharp discriminator of\\
(dominant) size, but a very poor discriminator of composition.''}

\hspace{68mm}{\sf ------ H.C. van de Hulst [1989]}
\vspace{2mm}

The most extensively studied dust property 
may be the interstellar extinction.
The main characteristics of the wavelength dependence 
of interstellar extinction -- ``interstellar extinction 
curve'' -- are well established: a slow and then increasingly 
rapid rise from the IR to the visual, an approach to levelling 
off in the near UV, a broad absorption feature at about 
$\lambda^{-1} \approx 4.6\umrev$ ($\lambda \approx 2175\Angstrom$)
and, after the drop-off, a ``final'' curving increase to 
as far as has been observed $\lambda^{-1} \approx 8\umrev$. 
We note here that the access to the UV was only made possible 
when observations could be made from space, first by rockets 
and then by satellites (OAO2 and IUE).

The optical/UV extinction curves show 
considerable variations which are correlated with 
different regions.\footnote{%
  Greenberg \& Chlewicki (1983) found that the strength of 
  the 2175$\Angstrom$ hump and the far-UV extinction can 
  vary both independently and with respect to the visual 
  extinction. This may imply that the 2175$\Angstrom$ hump
  and the far-UV extinction are produced by two different
  dust components.
  }
Cardelli, Clayton, \& Mathis (1989) found that the extinction 
curves over the wavelength range of 
$0.125\um \le \lambda \le 3.5\um$ can be fitted remarkably well
by an analytical formula involving only one free parameter: 
$\Rv\equiv A_V/E(B-V)$, the total-to-selective extinction ratio.
Values of $\Rv$ as small as 2.1 (the high latitude translucent 
molecular cloud HD 210121; Larson, Whittet, \& Hough 1996) 
and as large as 5.6 (the HD 36982 molecular cloud in 
the Orion nebula) have been observed in the Galactic regions.
More extreme extinction curves are reported for gravitational
lens galaxies: $\Rv=1.5$ for an elliptical galaxy at
a lens redshift $z_l=0.96$ 
and $\Rv=7.2$ for a spiral galaxy at $z_l=0.68$
(Falco et al.\ 1999). The Galactic mean extinction curve
is characterized by $\Rv\approx 3.1$. 
The optical/UV extinction curve and the value of $\Rv$ 
depend on the environment: lower-density regions have 
a smaller $\Rv$, a stronger 2175$\Angstrom$ hump and 
a steeper far-UV rise ($\lambda^{-1} > 4\umrev$); 
denser regions have a larger $\Rv$, a weaker 2175$\Angstrom$
hump and a flatter far-UV rise. 

The near-IR extinction curve ($0.9\um \le \lambda \le 3.5\um$)
can be fitted reasonably well by a power law 
$A(\lambda) \sim \lambda^{-1.7}$, showing little environmental 
variations. The extinction longward of 3$\um$ is not as well
determined as that of $\lambda \le 3\um$. Lutz et al.\ (1996)
derived the 2.5--9$\um$ extinction law toward the Galactic
Center (GC) based on the ISO observation of the hydrogen 
recombination lines. They found that the GC extinction law 
has a higher extinction level in the 3--8$\um$ range than
the standard Draine (1989a) IR extinction curve.\footnote{%
  The silicate core-organic refractory mantle model 
  (Greenberg 1989; Li \& Greenberg 1997) may provide 
  a better fit to the 3--8$\um$ extinction curve than 
  the silicate-graphite model (Mathis et al.\ 1977; 
  Draine \& Lee 1984) because the carbonaceous organic
  material is IR active at $\lambda > 5\um$ due to 
  ${\rm C=C}$, ${\rm C=O}$, ${\rm C-OH}$, ${\rm C\equiv N}$, 
  ${\rm C-NH_2}$ stretches; CH, OH, and NH$_2$ deformations,
  and H wagging (Greenberg et al.\ 1995).
  }
 
Existing grain models for the diffuse interstellar medium 
are mainly based on an analysis of extinction 
(Mathis, Rumpl, \& Nordsieck 1977; Greenberg 1978; 
Hong \& Greenberg 1980; Draine \& Lee 1984; 
Duley, Jones, \& Williams 1989; Mathis \& Whiffen 1989; 
Kim, Martin, \& Hendry 1994; Mathis 1996; Li \& Greenberg 1997; 
Zubko 1999a; Weingartner \& Draine 2001a). All models are 
successful in reproducing the observed extinction curve 
from the near IR to the far-UV. 
The silicate core-organic refractory mantle model
and the (modified) silicate/graphite model
are also able to fit the HD 210121 extinction curve which
has the lowest $\Rv$ value (Li \& Greenberg 1998d; 
Larson et al.\ 2000; Clayton et al.\ 2000; 
Weingartner \& Draine 2001a) and the Magellanic clouds 
extinction curves (Rodrigues et al.\ 1997; Zubko 1999b; 
Clayton et al.\ 2000; Weingartner \& Draine 2001a). 
The fact that so many different materials with such 
a wide range of optical properties could be used to 
explain the observed extinction curve indicates that
the interstellar extinction curve is quite insensitive
to the exact dust composition. What the extinction curve
does tell us is that interstellar grains span a size ranging
from $\sim 100$ angstrom to submicron (Draine 1995).

The general shape of the polarization curve is also well 
established. It rises from the IR, has a maximum 
somewhere in the visual (generally) and then decreases 
toward the UV, implying that the aligned nonspherical 
grains are typically submicron in size, and the 
very small grain component responsible for the far-UV 
extinction rise is either spherical or not aligned. 

The optical/UV polarization curve $P(\lambda)$ can 
also be fitted remarkably well by an empirical function
known as the ``Serkowski law'', involving the very same 
free parameter $\Rv$ as the Cardelli et al.\ (1989) extinction
functional form through $\lambdamax$,\footnote{%
  $\Rv \approx (5.6\pm 0.3)\lambdamax$ 
  ($\lambdamax$ is in micron; see Whittet 1992);
  i.e., the wavelength of maximum polarization $\lambdamax$ 
  shifts with $\Rv$ in the sense that it moves to longer 
  wavelengths as $\Rv$ increases which is the effect of 
  increasing the particle size.
  }
the wavelength where the maximum polarization $\Pmax$ occurs: 
$P(\lambda)/\Pmax = \exp [-K\ln^2(\lambda/\lambdamax)]$
(Serkowski 1973; Coyne, Gehrels, \& Serkowski 1974;
Wilking, Lebofsky, \& Rieke 1982). 
The width parameter $K$ is linearly correlated with
$\lambdamax$: $K\approx 1.66\,\lambdamax\,+\,0.01$ 
(Whittet et al.\ 1992 and references therein).\footnote{%
  The far-UV polarization observations only became available  
  in recent years as a consequence of the Wisconsin Ultraviolet 
  Photo-Polarimetry Experiment (WUPPE) (Clayton et al.\ 1992) 
  and the UV polarimetry of the Hubble Space Telescope
  (Somerville et al. 1993, 1994; Clayton et al. 1995). 
  It is found that 
  (1) lines of sight with $\lambdamax \geq 0.54\um$ 
  are consistent with the extrapolated Serkowski law; 
  (2) lines of sight with $\lambdamax \leq 0.53\um$ show 
  polarization in excess of the extrapolated Serkowski law; 
  (3) two lines of sights show a polarization feature which 
  seems to be associated with the $4.6\umrev$ extinction hump 
  (Clayton et al.\ 1992; Anderson et al.\ 1996; Wolff et al.\ 1997;
   Martin, Clayton, \& Wolff 1999).
  } 

The near IR ($1.64\um < \lambda < 5\um$) polarization 
has been found to be higher than that extrapolated from 
the Serkowski law (Martin et al.\ 1992). 
Martin \& Whittet (1990) and Martin et al.\ (1992) suggested
a power law $P(\lambda) \propto \lambda^{-\beta}$ 
for the near IR ($\lambda > 1.64\um$) where the power index 
$\beta$ is independent of $\lambda_{\rm max}$ and in the range of 
1.6 to 2.0, $\beta \simeq 1.8\pm 0.2$.\footnote{%
  Martin et al.\ (1999) proposed a more complicated formula
  -- the ``modified Serkowski law'' --
  to represent the observed interstellar polarization
  from the near IR to the far UV.  
  } 

The observed interstellar polarization curve has 
also been extensively modelled in terms of various 
dust models by various workers.
The silicate core-organic mantle model 
(Chlewicki \& Greenberg 1990; Li \& Greenberg 1997),
the silicate-graphite model (only silicate grains are assumed to
be efficiently aligned; Mathis 1986; Wolff, Clayton, \& Meade 1993; 
Kim \& Martin 1995, 1996), 
and the composite model (Mathis \& Whiffen 1989) 
are all successful in reproducing the mean interstellar 
polarization curve ($\lambdamax = 0.55\um$).  

However, the processes leading to the observed grain 
alignment are still not well established. A number 
of alignment mechanisms have been proposed.
The Davis-Greenstein paramagnetic dissipation mechanism 
(Davis \& Greenstein 1951) together with other co-operative 
effects such as suprathermal rotation (Purcell 1979),
superparamagnetic alignment (Jones \& Spitzer 1967; Mathis 1986), 
radiative torques on irregular grains due to anisotropic starlight 
(Draine \& Weingartner 1996, 1997) seems to be a plausible 
mechanism for dust in the diffuse ISM.
In dense molecular clouds, non-magnetic alignment mechanisms 
such as streaming of grains through gas (Gold 1952; Purcell 1969;
Lazarian 1994), through radiation (Harwit 1970), 
through ambipolar diffusion (Roberge 1996) have been studied.

The wavelength dependent albedo (the ratio of scattering cross 
section to extinction) measured from the diffuse Galactic 
light, reflecting the scattering properties of interstellar dust, 
provides another constraint on dust models. 
The silicate core-organic mantle model (Li \& Greenberg 1997) 
and the silicate/graphite-PAHs model (Li \& Draine 2001b) 
are shown to be in good agreement with the observationally
determined albedos, whereas the albedos of the composite 
grain model (Mathis 1996) are too low (Dwek 1997).

Scatterings of X-rays by interstellar dust have also been 
observed as evidenced by ``X-ray halos'' formed around an 
X-ray point source by small-angle scattering. The intensity 
and radial profile of the halo depends on the composition, 
size and morphology and the spatial distribution of 
the scattering dust particles (see Smith \& Dwek 1998 
and references therein). A recent study of the X-ray halo
around Nova Cygni 1992 by Witt, Smith, \& Dwek (2001) 
pointed to the requirement of large interstellar grains,
consistent with the recent Ulysses and Galileo detections
of interstellar dust entering our solar system 
(Gr\"{u}n et al.\ 1994; Frisch et al.\ 1999; Landgraf et al.\ 2000). 

\subsubsection{Spectroscopic Extinction and Polarization Features}
It is the extinction (absorption) and emission spectral 
lines instead of the overall shape of the extinction curve 
provides the most diagnostic information on the 
dust composition.

\begin{enumerate}

\item \underline{The 2175$\Angstrom$ Extinction Hump}

\vspace{2mm}
\hspace{-5mm}{\tt ``It is frustrating that almost 3 decades after 
its discovery,}\\
\vspace{-1mm} 
\hspace{-1mm}{\tt the identity of this (2175$\Angstrom$) 
feature remains uncertain!''}

\vspace{2mm}
\hspace{76mm}{\sf ------ B.T. Draine [1995]}
\vspace{1mm}

The strongest spectroscopic extinction feature is 
the 2175$\Angstrom$ hump. Observations show that its strength 
and width vary with environment while its peak position is 
quite invariant. Its carrier remains unidentified 37 years after 
its first detection (Stecher 1965). 
Many candidate materials, including graphite (Stecher \& Donn 1965), 
amorphous carbon (Bussoletti et al.\ 1987),
graphitized (dehydrogenated) hydrogenated amorphous carbon 
(Hecht 1986; Goebel 1987; Sorrell 1990; Mennella et al.\ 1996;
Blanco et al.\ 1999),\footnote{%
  Mennella et al.\ (1996) reported that a stable peak position 
  can be obtained by subjecting small hydrogenated amorphous carbon
  grains to UV radiation. However, the laboratory produced humps 
  are too wide and too weak with respect to the interstellar one.
  In a later paper (Mennella et al.\ 1998) they proposed that
  the 2175$\Angstrom$ carrier can be modelled as a linear combination
  of such materials exposed to different degrees of UV processing. 
  }
nano-sized hydrogenated amorphous carbon (Schnaiter et al.\ 1998),
quenched carbonaceous composite (QCC; Sakata et al.\ 1995), 
coals (Papoular et al.\ 1995), PAHs (Joblin et al.\ 1992; 
Duley \& Seahra 1998; Li \& Draine 2001b), 
and OH$^{-}$ ion in low-coordination sites on or within 
silicate grains (Duley, Jones \& Williams 1989)
have been proposed, while no single one is generally 
accepted (see Draine 1989b for a review). 

\hspace{4mm} Graphite was the earliest suggested and 
the widely adopted candidate in various dust models 
(Gilra 1972; Mathis et al.\ 1977; 
Hong \& Greenberg 1980; Draine \& Lee 1984; Dwek et al.\ 1997;
Will \& Aannestad 1999). 
However the hump peak position predicted from graphite particles 
is quite sensitive to the grain size, shape, and coatings 
(Gilra 1972; Greenberg \& Chlewicki 1983; Draine 1988; Draine \& 
Malhotra 1993) which is inconsistent with the observations. 
It was suggested that very small coated graphite particles 
($a \leq 0.006\um$) could broaden the hump while keeping the 
hump peak constant (Mathis 1994).\footnote{%
 Hecht (1981) investigated the effect of coatings on graphite 
 particles and concluded that small coatings could be present 
 on spherical $a \approx 200\Angstrom$ particles and still 
 cause the 2175$\Angstrom$ feature.
 }
However, this seems unlikely because the proposed particles 
are so small that temperature fluctuations will prevent them 
from acquiring a coating (Greenberg \& Hong 1974a; 
Aannestad \& Kenyon 1979).
Furthermore, it was noted by Greenberg \& Hong (1974b) 
that if the very small particles as well as the large 
particles accrete mantles the $\Rv$ value would {\it decrease} 
in molecular clouds, in contradiction with astronomical
observations. Rouleau, Henning, \& Stognienko (1997)
proposed that the combined effects of shape, clustering, 
and fine-tuning of the optical properties of graphite could 
account for the hump width variability.

Another negative for graphite, is that the dust grains 
in circumstellar envelopes around carbon stars which are 
the major sources of the carbon component of interstellar 
dust are in amorphous form rather than graphitic (Jura 1986). 
It is difficult to understand how the original amorphous 
carbonaceous grains blown out from the star envelopes 
are processed to be highly anisotropic and evolve to the
layer-lattice graphitic structures in interstellar space. 
Instead, it is more likely that the interstellar physical 
and chemical processes should make the carbonaceous grains 
even more highly disordered. 

\hspace{4mm} Recently, the PAH proposal is receiving 
increasing attention. Although a single PAH species 
often has some strong and narrow UV bands which are not 
observed (UV Atlas 1966), a cosmic mixture 
of many individual molecules, radicals, and ions, 
with a concentration of strong absorption features
in the 2000--2400$\Angstrom$ region,
may effectively produce the 2175$\Angstrom$ extinction feature
(Li \& Draine 2001b). This is supported by the correlation between 
the 2175$\Angstrom$ hump and the IRAS 12$\um$ emission (dominated 
by PAHs) found by Boulanger, Pr\'{e}vot, \& Gry (1994) 
in the Chamaeleon cloud which suggests a common carrier. 
Arguments against the PAHs proposal also exist 
(see Li \& Draine 2001b for references).

\hspace{4mm} So far only two lines of sight toward HD 147933 
and HD 197770 have a weak 2175$\Angstrom$ polarization 
feature detected (Clayton et al.\ 1992; Anderson et al.\ 1996; 
Wolff et al.\ 1997; Martin, Clayton, \& Wolff 1999). 
Even for these sightlines, the degree of alignment 
and/or polarizing ability of the carrier should be very small 
(if both the hump excess polarization and the hump extinction 
are produced by the same carrier); 
for example, along the line of sight to
HD 197770, the ratio of the excess polarization to 
the hump extinction is $\Phump/\Ahump \simeq 0.002$ 
while the polarization to extinction ratio in 
the visual is $P_V/A_V \simeq 0.025$, thus 
$(\Phump/\Ahump)/(P_V/A_V)$ is only $\sim 0.09$.
Therefore, it is reasonable to conclude that 
the 2175$\Angstrom$ carrier is either mainly spherical
or poorly aligned.  

\hspace{4mm} The 2175$\Angstrom$ hump polarization was predicted 
by Draine (1988) for aligned non-spherical graphite grains. 
Wolff et al.\ (1993) and Martin et al.\ (1995) further
show that the observed 2175$\Angstrom$ polarization feature 
toward HD 197770 can be well fitted with small aligned 
graphite disks.

\hspace{4mm} Except for the detection of scattering in the 
2175$\Angstrom$ hump in two reflection nebulae (Witt, Bohlin, 
\& Stecher 1986), the 2175$\Angstrom$ hump is thought to be
predominantly due to absorption, suggesting its carrier 
is sufficiently small to be in the Rayleigh limit.

\vspace{2mm}
\item \underline{The 9.7$\um$ and 18$\um$ (Silicate) Absorption
Features}

\vspace{1mm}
\hspace{21mm}{\tt ``In my opinion, the only secure identification}\\
\vspace{-1mm}
\hspace{22mm}{\tt is that of the 9.7$\um$ and 18$\um$ IR features.''}

\vspace{2mm}
\hspace{76mm}{\sf ------ B.T. Draine [1995]}
\vspace{1mm}

The strongest IR absorption features are the 9.7$\um$ and 18$\um$
bands. They are respectively ascribed to the Si-O stretch 
and O-Si-O bending modes in some form of silicate material, 
perhaps olivine Mg$_{2x}$Fe$_{2-2x}$SiO$_4$.
The shape of the interstellar silicate feature is broad and
featureless both of which suggest that the silicate 
is amorphous.\footnote{%
  Very recently, Li \& Draine (2001a) estimated that the 
  abundances of $a < 1\um$ crystalline silicate 
  grains in the diffuse ISM is $< 5\%$ of the solar Si
  abundance.
  }
The originating source of the interstellar silicates 
is in the atmospheres of cool evolved stars
of which the emission features often show 
a 9.7$\um$ feature consistent with amorphous silicates 
and sharper features arising from crystalline silicates
(Waters et al.\ 1996).\footnote{%
  Crystalline silicates have also been seen in six comets 
  (see Hanner 1999 for a summary), in dust disks around
  main-sequence stars (see Artymowicz 2000 for a summary), 
  young stellar objects (see Waelkens, Malfait, \& Waters 2000 
  for a summary), in interplanetary dust particles (IDPs) 
  (Bradley et al.\ 1999), and probably also in the Orion Nebula 
  (Cesarsky et al.\ 2000).
  } 
How can interstellar crystalline silicate particles 
become amorphous? This is a puzzle which has not yet been 
completely solved. A possible solution may be that the energetic
processes (e.g. shocks, grain-grain collision) operated on 
interstellar dust in the diffuse ISM have disordered the 
periodic lattice structures of crystalline silicates.
Another possible solution may lie in the fact that, 
according to Draine (1990), only a small fraction of 
interstellar dust is the original stardust, i.e., 
most of the dust mass in the ISM was condensed in the 
ISM, rather than in stellar outflows. The re-condensation
at low temperatures most likely leads to an amorphous rather 
than crystalline form.

\hspace{4mm} First detected in the Becklin-Neugebauer (BN) 
object in the OMC-1 Orion dense molecular cloud (Dyck et al.\ 1973),
the silicate polarization absorption feature in the 
10$\um$ region is found to be very common in heavily 
obscured sources; some sources also have the 18$\um$ 
O-Si-O polarization feature detected (see Aitken 1996;
Smith et al.\ 2000 for summaries). In most cases the
silicate polarization features are featureless, indicating
the amorphous nature of interstellar silicate material
(see Aitken 1996) except AFGL 2591, a molecular cloud 
surrounding a young stellar object, has an additional 
narrow polarization feature at 11.2$\um$, generally 
attributed to annealed silicates (Aitken et al.\ 1988;
Wright et al.\ 1999). 

\hspace{4mm} Reasonably good fits to the observed 10$\um$ 
Si-O polarization features can be obtained by elongated 
(bare or ice-coated) ``astronomical silicate'' grains 
(Draine \& Lee 1984; Lee \& Draine 1985; 
Hildebrand \& Dragovan 1995; Smith et al.\ 2000). 
High resolution observations of the BN 10$\um$ and 18$\um$ 
polarization features provided a challenge to the ``astronomical 
silicate'' model since this model failed to reproduce two of
the basic aspects of the observations: 
(1) the 10$\um$ feature was not broad enough 
and (2) the 18$\um$ feature was too low by a
factor of two relative to the 10$\um$ peak
(Aitken, Smith, \& Roche 1989).  
Attempts by Henning \& Stognienko (1993) in terms of
porous grains composed of ``astronomical silicates'',
carbon and vacuum were not successful.
In contrast, it is shown by Greenberg \& Li (1996) that
an excellent match to the BN 10$\um$ and 18$\um$ 
polarization features in shape, width, and in relative 
strength can be obtained by the silicate core-organic mantle 
model using the experimental optical constants of silicate
and organic refractory materials. It seems desirable to
re-investigate the 18$\um$ O-Si-O band strength of 
``astronomical silicates'' (Draine \& Lee 1984) by 
a combination of observational, experimental and modelling efforts. 

\hspace{4mm} In the mid-IR, interstellar grains of submicron 
size are in the Rayleigh limit. The scattering efficiency 
falls rapidly with increasing wavelength. 
Therefore the effects of scattering are expected to be 
negligible for the silicate extinction and polarization.

\vspace{2mm}
\item \underline{The 3.4$\um$ (Aliphatic Hydrocarbon) Absorption Feature}

\vspace{2mm}
\hspace{-12mm}{\tt ``When the 3.4$\um$ feature was detected in
VI Cyg \#12, the problem}\\
\vspace{-1mm}
\hspace{-10mm}{\tt of why no H$_2$O was observed would
appear to have been resolved.''}

\vspace{1mm}
\hspace{66mm}{\sf ------ J. Mayo Greenberg [1999]}
\vspace{2mm}

Another ubiquitous strong absorption band in the diffuse ISM
is the 3.4$\um$ feature. Since its first detection in
the Galactic Center toward Sgr A W by Willner et al.\ (1979) 
and IRS 7 by Wickramasinghe \& Allen (1980), 
it has now been widely seen in the Milky Way Galaxy and other 
galaxies (Butchart et al.\ 1986; Adamson, Whittet, \& Duley 1990; 
Sandford et al.\ 1991; Pendleton et al.\ 1994; Wright et al.\ 1996; 
Imanishi \& Dudley 2000; Imanishi 2000). Although it is generally 
accepted that this feature is due to the C-H stretching mode in 
saturated aliphatic hydrocarbons, the exact nature of this 
hydrocarbon material remains uncertain. Nearly two dozen different
candidates have been proposed over the past 20 years (see Pendleton
\& Allamandola 2002 for a review). The organic refractory residue,
synthesized from UV photoprocessing of interstellar ice mixtures, 
provides a perfect match, better than any other hydrocarbon
analogs, to the observed 3.4$\um$ band, 
including the 3.42$\um$, 3.48$\um$, and 3.51$\um$ subfeatures 
(Greenberg et al.\ 1995).\footnote{%
  Pendleton \& Allamandola (2002) questioned the robustness
  of the fit by the organic residue since the broad 5.5--10$\um$
  band seen in the organic residue spectrum was not observed
  in astronomical spectra. We note that this band, largely 
  attributed to the combined features of 
  the C=O, C-OH, C$\equiv$N, C-NH$_2$, OH, and NH$_2$ 
  stretches, bendings, and deformations, will become weaker 
  if the organics are subject to further UV photoprocessing 
  which will result in photodissociation and depletion of 
  H, O, N elements. The organic residue samples presented
  in Greenberg et al.\ (1995) were processed at most to a degree 
  resembling one cycle (from molecular clouds to diffuse clouds).
  According to the cyclic evolution model, interstellar grains will 
  undergo $\sim 50$ cycles before they are consumed by star formation
  or becomes a part of a comet (Greenberg \& Li 1999a). 
  }
But, at this moment, we are not at a position to rule out other 
dust sources as the interstellar 3.4$\um$ feature carrier.
This feature has also been detected in a carbon-rich 
protoplanetary nebula CRL 618 (Lequeux \& Jourdain de Muizon 1990; 
Chiar et al.\ 1998) with close resemblance to the interstellar 
feature. However, after ejection into interstellar space, 
the survival of this dust in the diffuse ISM is questionable 
(see Draine 1990). 
 
\hspace{4mm} The 3.4$\um$ feature consists of three subfeatures at 
${\rm 2955\,cm^{-1}}$ (3.385\\$\mu$m), 
${\rm 2925\,cm^{-1}}$ (3.420$\um$), 
and ${\rm 2870\,cm^{-1}}$ (3.485$\um$)
corresponding to the symmetric and asymmetric C-H stretches 
in CH$_3$ and CH$_2$ groups in aliphatic hydrocarbons which 
must be interacting with other chemical groups.
The amount of carbonaceous material responsible for the  
3.4$\um$ feature is strongly dependent on the nature 
of the chemical groups attached to the aliphatic carbons. 
For example, each carbonyl (C=O) group reduces its corresponding 
C-H stretch strength by a factor of $\sim 10$ (Wexler 1967). 
Furthermore, not every carbon is attached to a hydrogen as in 
saturated compounds. Aromatic hydrocarbons do not even contribute 
to the 3.4$\um$ feature although they do absorb nearby at 
$\approx 3.28\um$. The fact that the 3.4$\um$ absorption 
is not observed in molecular cloud may possibly be attributed to 
dehydrogenation or oxidation (formation of carbonyl) of the 
organic refractory mantle by accretion and photoprocessing in 
the dense molecular cloud medium (see Greenberg \& Li 1999a)
-- the former reducing the absolute number of CH stretches, 
the latter reducing the CH stretch strength by a factor of 10 
(Wexler 1967) -- the 3.4$\um$ feature would be reduced 
{\it per unit mass} in molecular clouds. 

\hspace{4mm} Very recently, Gibb \& Whittet (2002) 
reported the discovery of a 6.0$\um$ feature 
in dense clouds attributed to the organic refractory. 
They found that its strength is correlated with the 
4.62$\um$ OCN$^{-}$ (XCN) feature which is considered 
to be a diagnostic of energetic processing.

\hspace{4mm} Attempts to measure the polarization of the 3.4$\um$ 
absorption feature ($P_{\rm C-H}^{\rm IRS7-obs}$) was recently 
made by Adamson et al.\ (1999) toward the Galactic Center source 
IRS 7. They found that this feature was essentially unpolarized. 
Since no spectropolarimetric observation of the 10$\um$ silicate 
absorption feature ($P_{\rm sil}^{\rm IRS7}$) has yet been 
carried out for IRS 7, they estimated $P_{\rm sil}^{\rm IRS7}$
from the 10$\um$ silicate optical depth $\tau_{\rm sil}^{\rm IRS7}$, 
assuming the IRS 7 silicate feature is polarized to the same degree as 
the IRS 3 silicate feature; 
i.e., $P_{\rm sil}^{\rm IRS7}/\tau_{\rm sil}^{\rm IRS7} 
= P_{\rm sil}^{\rm IRS3}/\tau_{\rm sil}^{\rm IRS3}$
where $\tau_{\rm sil}^{\rm IRS7}$, $\tau_{\rm sil}^{\rm IRS3}$ 
and $P_{\rm sil}^{\rm IRS3}$ were known.   
Assuming the IRS 7 aliphatic carbon (the 3.4$\um$ carrier) is 
aligned to the same degree as the silicate dust, they expected 
the 3.4$\um$ polarization to be $P_{\rm C-H}^{\rm IRS7-mod}
= P_{\rm sil}^{\rm IRS7}/\tau_{\rm sil}^{\rm IRS7} 
\times \tau_{\rm C-H}^{\rm IRS7}$. They found 
$P_{\rm C-H}^{\rm IRS7-obs} \ll P_{\rm C-H}^{\rm IRS7-mod}$
(Adamson et al.\ 1999). Therefore, they concluded that 
the aliphatic carbon dust is not in the form of 
a mantle on the silicate dust as suggested by 
the core-mantle models (Li \& Greenberg 1997; 
Jones, Duley, \& Williams 1990).\footnote{%
 We note that the observed correlation between 
 the 10$\um$ silicate and the 3.4$\um$ C-H hydrocarbon 
 optical depths (Sandford, Pendleton, \& Allamandola 1995) 
 is consistent with the core-mantle scenario.  
 }
We note that the two key assumptions on which
their conclusion relies are questionable:
(1)  $P_{\rm sil}^{\rm IRS7}/\tau_{\rm sil}^{\rm IRS7} 
= P_{\rm sil}^{\rm IRS3}/\tau_{\rm sil}^{\rm IRS3}$;
(2) $P_{\rm C-H}^{\rm IRS7}/\tau_{\rm C-H}^{\rm IRS7} 
= P_{\rm sil}^{\rm IRS7}/\tau_{\rm sil}^{\rm IRS7}$
(see Li \& Greenberg 2002 for details).
We urgently need spectropolarimetric observations of IRS 7. 

\hspace{4mm}Hough et al.\ (1996) reported the 
detection of a weak 3.47$\um$ polarization feature 
in BN, attributed to carbonaceous materials with 
diamond-like structure, originally proposed by 
Allamandola et al.\ (1992) based on the 3.47$\um$ 
absorption spectra of protostars.\footnote{%
  Hydrogenated nanodiamonds were identified in
  the circumstellar dust envelopes surrounding
  two Herbig Ae/Be stars HD 97048 and Elias 1
  revealed by their 3.43$\um$ and 3.53$\um$ emission
  features (Guillois, Ledoux, \& Reynaud 1999;
  van Kerckhoven, Tielens, \& Waelkens 2002). 
  Interstellar diamonds were first proposed as
  a dust component by Saslaw \& Gaustad (1969) 
  and first discovered in meteorites by 
  Lewis, Anders, \& Draine (1989). 
  The fact that the 3.43$\um$ and 3.53$\um$ features 
  are not observed in the ISM led Tielens et al.\ (1999)
  to infer an upper limit of $\leq 0.1\ppm$ for
  hydrogenated interstellar nanodiamonds.  
  If interstellar diamonds are not hydrogenated,
  they could be much more abundant (van Kerckhoven et al.\ 2002). 
  Analysis of the interstellar extinction observations show 
  that up to 10\% of the interstellar carbon can be locked up 
  in diamond and escape detection (Lewis et al.\ 1989). 
  }

\vspace{2mm}
\item \underline{The 3.3$\um$ and 6.2$\um$ (PAH) Absorption Features}

Recently, two weak narrow absorption features at 3.3$\um$ 
and 6.2$\um$ were detected. The 3.3$\um$ feature has been seen in 
the Galactic Center source GCS 3 (Chiar et al.\ 2000) 
and in some heavily extincted molecular cloud sight lines 
(Sellgren et al.\ 1995; Brooke, Sellgren, \& Geballe 1999).
The 6.2$\um$ feature, about 10 times stronger, 
has been detected in several objects including both local 
sources and Galactic Center sources (Schutte et al.\ 1998;
Chiar et al.\ 2000). They were attributed to aromatic hydrocarbons
(Schutte et al.\ 1998; Chiar et al.\ 2000). The theoretical
3.3$\um$ and 6.2$\um$ absorption feature strengths 
(in terms of integrated optical depths) predicted from 
the astronomical PAH model are consistent with observations 
(Li \& Draine 2001b).
Note the 7.7, 8.6, and 11.3$\um$ PAH features are hidden 
by the much stronger 9.7$\um$ silicate feature, and therefore 
will be difficult to observe as absorption features.

\hspace{4mm} The aromatic absorption bands allow one to place 
constraints on the PAH abundance if the PAH band strengths 
are known. But one should keep in mind that the strengths 
of these PAH absorption bands could vary with physical 
conditions due to changes in the PAH ionization fraction. 
In regions with increased PAH ionization fraction, 
the 6.2$\um$ absorption feature would be strengthened
and the 3.3$\um$ feature would be weakened (see Li \& Draine 2001b). 

\hspace{4mm} Although it was theoretically predicted that 
the PAH IR emission features can be linearly polarized 
(L\'{e}ger 1988), no polarization has been detected yet 
(Sellgren, Rouan, \& L\'{e}ger 1988). 

\vspace{2mm}
\item \underline{The Diffuse Interstellar Bands}
\vspace{1mm}

In 1922, Heger observed two broad absorption features 
centering at 5780$\Angstrom$ and 5797$\Angstrom$,
conspicuously broader than atomic interstellar absorption 
lines. This marked the birth of a long standing astrophysical 
mystery -- the diffuse interstellar bands (DIBs). 
But not until the work of Merrill (1934) were 
the interstellar nature of these absorption features 
established. So far, over 300 DIBs have been detected
from the near IR to the near UV. Despite $\sim 80$ years' 
efforts, no definite identification (including the recent
neutral/charged PAHs [Salama \& Allamandola 1992], 
C$_{\rm 60}^{+}$ [Foing \& Ehrenfreund 1994], 
and C$_{7}^{-}$ [Tulej et al.\ 1998] proposals) of 
the carrier(s) of DIBs has been found. We refer the 
readers to the two extensive reviews of Krelowski (1999, 2002).  

\hspace{4mm} No polarization has been detected for the DIBs 
(Martin \& Angel 1974, 1975; Fahlman \& Walker 1975;
Adamson \& Whittet 1992, 1995; 
see Somerville 1996 for a review).

\vspace{2mm}
\item \underline{The Ice Absorption Features}
\vspace{1mm}

The formation of an icy mantle through accretion of molecules 
on interstellar dust is expected to take place in dense clouds 
(e.g., the accretion timescale is only $\sim 10^5$ yrs for clouds 
of densities $n_{\rm H} = 10^3-10^5\cm^{-3}$; Schutte 1996). 
The detection of various ice IR absorption features 
(e.g., H$_2$O [3.05, 6.0$\um$], CO [4.67$\um$], 
CO$_2$ [4.27, 15.2$\um$], CH$_3$OH [3.54, 9.75$\um$], 
NH$_3$ [2.97$\um$], CH$_4$ [7.68$\um$],
H$_2$CO [5.81$\um$], OCN$^{-}$ [4.62$\um$];
see Ehrenfreund \& Schutte 2000 for a review)
have demonstrated the presence of icy mantles in 
dark clouds (usually with an visual extinction $>3$ magnitudes;
see Whittet et al.\ 2001 and references therein). 
In comparison with that for a gas phase sample,
the IR spectrum for the same sample in the solid phase
is broadened, smoothed, and shifted in wavelength
due to the interactions between the vibrating molecule
with the surroundings, the suppression of molecular
rotation in ices at low temperatures, and the irregular
nature of the structure of (amorphous) solids 
(see Tielens \& Allamandola 1987).    

\hspace{4mm} Note not only are the relative proportions of ice species
variable in different regions but also the presence and
absence of some species. In almost all cases, however, 
water is the dominant component.
An important variability is in the layering of 
the various molecular components. 
Of particular note is the fact that the CO molecular spectrum 
is seen to indicate that it occurs sometimes embedded in the H$_2$O 
(a polar matrix) and sometimes not. This tells a story about 
how the mantles form. As was first noted by van de Hulst, 
the presence of surface reactions leads to the reduced species 
H$_2$O, CH$_4$, NH$_3$. Since we now know that CO is an abundant 
species as a gas phase molecule, we expect to find it accreted along
with these reduced species --- at least initially.

\hspace{4mm} The two approaches to understanding how the grain mantles 
evolve are: (1) the laboratory studies of icy mixtures, 
their modification by UV photoprocessing and by heating; 
(2) theoretical studies combining gas phase chemistry with 
dust accretion and dust chemistry. In the laboratory one
creates a cold surface (10K) on which various simple 
molecules are slowly deposited in various proportions. 
The processing of these mixtures by UV photons 
and by temperature variation is studied by IR spectroscopy. 
This analog of interstellar dust mantles is used to provide 
a data base for comparison with the observations
(see Schutte 1999 for a review).  

\hspace{4mm} The 3.1$\um$ ice polarization has been detected in various
molecular cloud sources (see Aitken 1996 and references therein). 
The detection of the 4.67$\um$ CO and 4.62$\um$ OCN$^{-}$
polarization was recently reported by Chrysostomou et al.\ (1996).
The BN ice polarization feature was well fitted by ice-coated 
grains (Lee \& Draine 1985), suggesting a core-mantle grain 
morphology. However, the AFGL 2591 molecular cloud shows no
evidence for ice polarization (Dyck \& Lonsdale 1980; Kobayashi
et al.\ 1981) while having distinct ice extinction and silicate
polarization (Aitken et al.\ 1988). 
Perhaps only the hot, partly annealed silicate grains close
to the forming-star are aligned (say, by streaming of ambipolar 
diffusion) while the ice-coated cool grains in the outer envelope
of the cloud are poorly aligned.  

\end{enumerate}

\subsection{Dust Emission}

\subsubsection{Dust Luminescence: The ``Extended Red Emission''}

\vspace{2mm}
\hspace{16mm}{\tt ``The ERE has become an important observational
aspect of interstellar grains that future models need to reproduce.''}

\hspace{82mm}{\sf ------ A.N. Witt [2000]}
\vspace{2mm}

First detected in the Red Rectangle (Schmidt, Cohen, \& Margon 
1980), ``extended red emission'' (ERE) from interstellar dust
consists of a broad, featureless emission band 
between $\sim$5400$\Angstrom$ and 9000$\Angstrom$,
peaking at $6100\ltsim \lambda_{\rm p} \ltsim 8200\Angstrom$, and
with a width $600\Angstrom\ltsim {\rm FWHM}\ltsim 1000\Angstrom$.
The ERE has been seen in a wide variety of dusty environments: 
the diffuse ISM of our Galaxy, reflection nebulae, planetary nebulae, 
HII regions, and other galaxies (see Witt, Gordon, \& Furton 1998 
for a summary). The ERE is generally attributed to 
photoluminescence (PL) by some component of interstellar dust, 
powered by UV/visible photons. The photon conversion efficiency of
the diffuse ISM has been determined to be near 
($10\pm 3$)\% (Gordon et al.\ 1998; Szomoru \& Guhathakurta 1998) assuming
that all UV/visible photons absorbed by interstellar grains are absorbed
by the ERE carrier.
The actual photoluminescence efficiency of the ERE carrier
must exceed $\sim 10\%$, since the ERE carrier
cannot be the only UV/visible photon absorber.

Various forms of carbonaceous materials -- 
HAC (Duley 1985; Witt \& Schild 1988), 
PAHs (d'Hendecourt et al.\ 1986),
QCC (Sakata et al.\ 1992), 
C$_{60}$ (Webster 1993), 
coal (Papoular et al.\ 1996), 
PAH clusters (Allamandola, private communication), 
carbon nanoparticles (Seahra \& Duley 1999),
and crystalline silicon nanoparticles (Witt et al.\ 1998;
Ledoux et al.\ 1998) -- have been proposed as carriers of ERE. 
However, most candidates appear to be unable 
to simultaneously match the observed ERE spectra 
and the required PL efficiency (see Witt et al.\ 1998 for details).

Although high photoluminescence efficiencies can be 
obtained by PAHs, the lack of spatial correlation 
between the ERE and the PAH IR emission bands 
in the compact HII region Sh 152 (Darbon et al.\ 2000), 
the Orion Nebula (Perrin \& Sivan 1992),
and the Red Rectangle (Kerr et al.\ 1999),
and the detection of ERE in the Bubble Nebula 
where no PAH emission has been detected (Sivan \& Perrin 1993)
seem against PAHs as ERE carriers.

Seahra \& Duley (1999) argued that small carbon 
clusters were able to meet both the ERE profile and 
the PL efficiency requirements. 
However, this hypothesis appears to be ruled out by
non-detection in NGC 7023 of the 1$\um$ ERE peak 
(Gordon et al.\ 2000)
predicted by the carbon nanoparticle model. 

Witt et al.\ (1998) and Ledoux et al.\ (1998) 
suggested crystalline silicon nanoparticles (SNPs) with
15\AA\ -- 50\AA\ diameters as the carrier on the 
basis of experimental data showing that SNPs could provide a close 
match to the observed ERE spectra and satisfy the quantum efficiency 
requirement. Smith \& Witt (2001) have further developed the SNP
model for the ERE, concluding that the observed ERE in the diffuse ISM
can be explained with Si/H = 6~ppm in SiO$_2$-coated SNPs with Si core
radii $a\approx17.5\Angstrom$.

Li \& Draine (2002b) calculated the thermal emission 
expected from such particles, both in a reflection nebula 
such as NGC 2023 and in the diffuse ISM. 
They found that Si/SiO$_2$ SNPs (both neutral and charged) 
would produce a strong emission feature at 20$\um$. 
The observational upper limit on the 20$\um$ feature
in NGC 2023 imposes an upper limit of $< 0.2$ppm Si
in Si/SiO$_2$ SNPs.
The ERE emissivity of the diffuse ISM appears to 
require $>15\ppm$ ($\gtsim$42\% of solar Si abundance) 
in Si/SiO$_2$ SNPs. 
In comparison with the predicted IR emission spectra,
they found that the DIRBE ({\it Diffuse Infrared 
Background Experiment}) photometry appears to rule out 
such high abundances of free-flying SNPs in the diffuse ISM. 
Therefore they concluded that if the ERE is due to SNPs, 
they must be either in clusters or attached to larger grains.
Future observations by SIRTF will be even more sensitive to the
presence of free-flying SNPs.

\subsubsection{Dust Temperatures and IR Emission}
The temperatures of interstellar dust particles 
depend on their optical properties and sizes 
(i.e., on the way they absorb and emit radiation) 
as well as on the interstellar radiation field (ISRF).\footnote{%
  Originally, the ISRF was represented by a $10^4\K$ 
  black-body radiation diluted by a factor 
  of $10^{-14}$ (Eddington 1926) which is undoubtedly 
  too crude but serves as a simple and adequate 
  approximation for some purposes. Many attempts 
  have been made to obtain a more reasonable determination 
  of the ISRF either on the basis of direct measurements 
  of the UV radiation from the sky or by calculating 
  the radiation of hot stars using model atmospheres. 
  Van Dishoeck (1994) has summarized the typical ISRF 
  estimates (Habing 1968; Draine 1978; Gondhalekar et al.\ 1980;
  Mathis, Mezger, \& Panagia 1983). As illustrated in 
  Fig.\,2 of van Dishoeck (1994), the various estimates 
  agree within factors of two. The latest work on the
  local far-UV ISRF by Parravano, Hollenbach, \& Mckee
  (2002) led to a value quite close to Draine (1978).
  } 
Most of the visible and UV radiation in galaxies from stars passes 
through clouds of particles and heats them. This heating leads to 
reradiation at much longer wavelengths extending to the millimeter. 
On the average, in spiral galaxies, $\sim 1/4 - 1/3$ of the total 
stellar radiation is converted into dust emission (Cox \& Mezger 1989;
Calzetti 2001). The converted radiation is a probe of the particles 
and the physical environments in which they find themselves. 

There is a long history in the study of grain temperature 
(and emission) since Eddington's demonstration of a 3.2$\K$ 
black body equilibrium dust temperature assuming 
a $10^4\K$ interstellar radiation field diluted by a factor of 
$10^{-14}$ (Eddington 1926). Van de Hulst (1949) was the first 
to provide a realistic dust model temperature, $\sim 15\K$ 
for dielectric particles. A subsequent extensive investigation 
was made by Greenberg (1968, 1971) where the temperatures were
calculated for various grain types in regions of various 
radiation fields. The first step to study the shape effects 
on dust temperatures was taken by Greenberg \& Shah (1971).
They found that the temperatures of non-spherical dielectric 
grains are generally lower than those of equivalent spheres, 
but insensitive to modest shape variations. 
Later efforts made by Chlewicki (1987) and
Voshchinnikov, Semenov, \& Henning (1999) essentially
confirmed the results of Greenberg \& Shah (1971).

The advent of the IRAS, COBE, and ISO space IR measurements 
provided powerful information regarding the far-IR emission 
of the large particles (the so-called ``cold dust''). 
The ``cold dust'' problem has received much attention since 
it plays an important role in many astrophysical subjects; 
for example, the presence of ``cold dust'' would change the current 
concept on the morphology and physics of galaxies (Block 1996). 

The presence of a population of ultrasmall grains
was known long before the IR era.
Forty-six years ago, Platt (1956) proposed that very small grains 
or large molecules with radii $\simlt 10$\AA\ may be present in 
interstellar space. Donn (1968) further proposed that 
polycyclic aromatic hydrocarbon-like ``Platt particles'', 
may be responsible for the UV interstellar extinction. 

These very small grains -- consisting of tens to hundreds 
of atoms -- are small enough that the time-averaged 
vibrational energy $\langle E\rangle$ is smaller than 
or comparable to the energy of the starlight photons
which heat the grains. Stochastic heating by absorption 
of starlight therefore results in transient ``temperature 
spikes'', during which much of the energy deposited
by the starlight photon is reradiated in the IR.
The idea of transient heating of very small grains was 
first introduced by Greenberg (1968). Since then,
there have been a number of studies on this topic
(see Draine \& Li 2001 and references therein).

Since the 1980s, an important new window on the 
``very small grain component'' has been opened by
IR observations. The near-IR continuum emission 
of reflection nebulae (Sellgren, Werner, \& Dinerstein 1983) 
and the 12 and 25$\um$ ``cirrus'' emission detected by IRAS 
(Boulanger \& P\'{e}rault 1988) explicitly indicated the 
presence of a very small interstellar dust component
since large grains (with radii $\sim 0.1\mu$m) heated by 
diffuse starlight emit negligibly at such short wavelengths, 
whereas very small grains (with radii $\simlt 0.01\mu$m) 
can be transiently heated to very high temperatures 
($\simgt 1000$\ K depending on grain size, composition, 
and photon energy). Subsequent measurements by the DIRBE 
instrument on the COBE satellite confirmed this and 
detected additional broadband emission 
at 3.5 and 4.9$\um$ (Arendt et al.\ 1998). 

More recently, spectrometers aboard 
the {\it Infrared Telescope in Space} 
(IRTS; Onaka et al.\ 1996; Tanaka et al.\ 1996) 
and ISO (Mattila et al.\ 1996) 
have shown that the diffuse ISM radiates strongly in 
emission features at 3.3, 6.2, 7.7, 8.6, and 11.3$\um$. 

\vspace{2mm}
\hspace{58mm}{\tt ``PAHs, they are everywhere!''}

\hspace{70mm}{\sf ------ L.J. Allamandola [1996]}
\vspace{2mm}

These emission features, first seen in the spectrum of the 
planetary nebulae NGC 7027 and BD+30$^{\rm o}$3639 
(Gillett, Forrest, \& Merrill 1973), have been observed 
in a wide range of astronomical environments including 
planetary nebulae, protoplanetary nebulae, 
reflection nebulae, HII regions, circumstellar envelopes, and
external galaxies (see Tielens et al.\ 1999 for a review
for Galactic sources and Helou 2000 for extragalactic sources). 
Often referred to as ``unidentified infrared'' (UIR) bands, 
these emission features are now usually attributed to 
PAHs which are vibrationally excited upon absorption of 
a single UV/visible photon (L\'{e}ger \& Puget 1984; 
Allamandola, Tielens, \& Barker 1985) 
although other carriers have also been proposed such as 
HAC (Duley \& Williams 1981; Borghesi, Bussoletti, \& Colangeli 1987; 
Jones, Duley, \& Williams 1990), QCC (Sakata et al.\ 1990), 
coal (Papoular et al.\ 1993), fullerenes (Webster 1993), 
and interstellar nanodiamonds with $sp^3$ surface atoms 
reconstructed to $sp^2$ hybridization (Jones \& d'Hendecourt 2000).

The emission mechanism proposed for the UIR bands
-- UV excitation of gas-phase PAHs followed by 
internal conversion and IR fluorescence\footnote{%
  PAHs are actually excited by photons of a wide
  range of wavelengths (Li \& Draine 2002c).
  }
-- is supported by laboratory measurements of the IR {\it emission} 
spectra of gas-phase PAH molecules (Cherchneff \& Barker 1989; 
Brenner \& Barker 1989; Kurtz 1992; Cook et al.\ 1998)
and by theoretical investigations of the heating and cooling processes 
of PAHs in interstellar space (Allamandola, Tielens, \& Barker 1989; 
Barker \& Cherchneff 1989; d'Hendecourt et al.\ 1989;
Draine \& Li 2001a).

The near-IR (1--5$\um$), mid-IR (5--12$\um$) emission spectrum 
along with the far-IR ($>$12$\um$) continuum emission of the 
diffuse Galactic medium yields further insights into the 
composition and physical nature of interstellar dust; 
in particular, the PAH emission features allow us to place 
constraints on the size distribution of the very small dust 
component.    

Attempts to model the IR emission of interstellar dust have
been made by various workers. Following the initial
detection of 60 and 100$\um$ cirrus emission (Low et al.\ 1984),
Draine \& Anderson (1985) calculated the IR emission from 
a graphite/silicate grain model with grains as small as 3\AA\
and argued that the 60 and 100$\um$ emission could be accounted
for. When further processing of the IRAS data revealed 
stronger-than-expected 12 and 25$\um$ emission from interstellar
clouds (Boulanger, Baud, \& van Albada 1985), Weiland et al.\ (1986) 
showed that this emission could be explained if very large numbers
of 3--10\AA\ grains were present. A step forward was taken by 
D\'{e}sert, Boulanger, \& Puget (1990), Siebenmorgen \& Kr\"{u}gel 
(1992), Schutte, Tielens, \& Allamandola (1993), and Dwek et al.\ (1997) 
by including PAHs as an essential grain component. 
Early studies were limited to the IRAS observation 
in four broad photometric bands, but Dwek et al.\ (1997) were able 
to use DIRBE and FIRAS data.

In recent years, there has been considerable progress 
in both experimental measurements and quantum chemical 
calculations of the optical properties of PAHs 
(Allamandola, Hudgins, \& Sandford 1999;
Langhoff 1996; and references therein).
There is also an improved understanding of the heat capacities 
of dust candidate materials (Draine \& Li 2001)
and the stochastic heating of very small grains 
(Barker \& Cherchneff 1989; d'Hendecourt et al.\ 1989; 
Draine \& Li 2001),
the interstellar dust size distributions 
(Weingartner \& Draine 2001a), and the grain charging 
processes (Weingartner \& Draine 2001b).

Li \& Draine (2001b) have made use of these advances 
to model the full emission spectrum, from near-IR to submillimeter, 
of dust in the diffuse ISM. The model consists of 
a mixture of amorphous silicate grains and carbonaceous grains,
each with a wide size distribution ranging 
from molecules containing tens of atoms
to large grains $\gtsim 1\um$ in diameter.
The carbonaceous grains are assumed to have
PAH-like properties at very small sizes, and graphitic properties 
for radii $a \gtsim 50$\AA.
On the basis of recent laboratory studies and 
guided by astronomical observations, they have constructed 
``astronomical'' absorption cross sections for use in modelling 
neutral and ionized PAHs from the far UV to the far IR.
Using realistic heat capacities (for calculating energy distribution 
functions for small grains undergoing ``temperature spikes''),
realistic optical properties,
and a grain size distribution consistent with the observed
interstellar extinction (Weingartner \& Draine 2001a),
Li \& Draine (2001b) were able to reproduce the near-IR to submillimeter 
emission spectrum of the diffuse ISM, including the PAH emission 
features at 3.3, 6.2, 7.7, 8.6, and 11.3$\micron$.

The silicate/graphite-PAH model has been shown also 
applicable to the Small Magellanic Cloud (Li \& Draine 2002a;
Weingartner \& Draine 2001a).

Li \& Draine (2002c) have also modelled the excitation 
of PAH molecules in UV-poor regions. 
It was shown that the astronomical PAH model provides
a satisfactory fit to the UIR spectrum of vdB 133, 
a reflection nebulae with the lowest ratio of UV to total
radiation among reflection nebulae with detected UIR band emission
(Uchida, Sellgren, \& Werner 1998).  

\subsubsection{Microwave Emission: Spinning Dust Grains} 
A number of physical processes including collisions with 
neutral atoms and ions, plasma drag, absorption and emission 
of photons can drive ultrasmall grains to rapidly rotate
(Draine \& Lazarian 1998a, 1998b; Draine \& Li 2002). 
The electric dipole emission from these spinning dust grains
was shown to be able to account for the 10--100\,GHz ``anomalous''
Galactic background component (See Draine \& Lazarian 1998b
and references therein).

\subsection{Interstellar Depletions}

\vspace{2mm}
\hspace{54mm}{\tt ``Where have all these atoms gone?''}

\hspace{68mm}{\sf ------ J. Mayo Greenberg [1963]}
\vspace{2mm}

Derivations of the relative abundances of the elements 
in our Galaxy are one of the principal needs for understanding 
the chemical evolution in interstellar space -- and ultimately 
its memory in comets. A major factor in developing consistent 
dust models was the observation of the ``depletion'' in 
low density clouds (atoms locked up in grains are ``depleted''
from the gas phase) using the UV absorption line spectroscopy
as a probe of the gas-phase abundances and assuming a reference
abundance (abundances of atoms both in gas and in dust).
 
The deduced possible dust composition was initially only 
constrained to the extent that silicates alone could not 
be responsible for the interstellar extinction (Greenberg 1974). 
But in recent years, the problem of grain modelling has been 
exacerbated by the apparent decrease of the available condensible 
atoms (O, C, N, Si, Mg, Fe) by about 30\% (Snow \& Witt 1996)
since the solar system was born.\footnote{%
  But see also Sofia \& Meyer (2001), who argue that 
  interstellar abundances are approximately solar.
  It is also possible that the solar system formed 
  out of material with a higher metallicity than 
  the average ISM at that time (there is evidence 
  that stars with planets have higher metallicities 
  than equal age stars without planets).
  So the metallicity of the ISM may not have declined 
  since the solar system was formed 
  (Draine, private communication). 
  }
This implies that the heavy elements are being consumed 
more than they are being created. However, if one goes back 
far enough in time, there were no condensible atoms because 
their initial production must follow the birth of stars. 
This brings us to the cosmological question of what do 
high-$z$ galaxies look like and when and how was dust
first found in them?

\subsection{Interstellar Dust in the Solar System}
Interstellar grains have been found in primitive
meteorites and in interplanetary dust particles based on 
the analysis of isotopic anomalies (see Kerridge 1999
and Bradley 1999 for recent reviews). 
Most presolar grains identified to date are 
carbonaceous: diamonds, SiC, and graphite (very small
TiC, ZrC, and MoC grains have also been found as
inclusions in SiC and graphite grains);
also identified are oxides such as corundum (Al$_2$O$_3$)
and silicon nitride (Si$_3$N$_4$).
One should keep in mind that much of the less refractory 
dust incorporated into meteorites is lost during
the chemical processing used to extract the refractory
grains from meteorites. Therefore, the extracted 
presolar grains are compositionally not representative 
of the bulk interstellar dust;\footnote{%
  We have already seen in Footnote-22 that diamond
  is not a major interstellar dust component. 
  Whittet, Duley, \& Martin (1990) found that
  the abundance of Si in SiC dust in the diffuse ISM 
  is at most 5\% of that in silicates.}  
for example, the procedures 
used to isolate interstellar grains in meteorites are designed
to deliberately destroy silicate material which
constitutes the bulk of the host meteorite (see Draine 1994). 

The solar system is surrounded by the local interstellar
cloud with a density $n_{\rm H} \approx 0.3\cm^{-3}$ and 
moving past the Sun with a velocity $\approx 26\km\s^{-1}$
(Lallement et al.\ 1994). Interstellar grains embedded in 
the local cloud with sufficiently low charge-to-mass ratios 
can penetrate the heliopause and enter the solar system on 
hyperbolic orbits (small, charged grains are deflected from
the heliosphere; Linde \& Gombosi 2000). The interplanetary 
spacecraft Ulysses and Galileo have detected over 600 grains 
flowing into the solar system and determined their speed, 
direction, and mass and therefore the mass flux (but not 
chemical composition since the grains were destroyed by the
detection technique; Gr\"{u}n et al.\ 1993, 1994; Frisch et al.\ 1999; 
Landgraf et al.\ 2000). Including the mass of the large population 
of interstellar micrometeorites entering the Earth's atmosphere 
(Taylor et al.\ 1996; Baggaley 2000), Frisch et al.\ (1999)
found that the total dust-to-gas mass ratio in the local 
interstellar cloud is about twice the canonical
value determined from the interstellar extinction. 
They also found that there is a substantial amount of mass 
in large grains of $\sim 1\um$ in size which is difficult to
reconcile with the interstellar extinction and interstellar
elemental abundances. 

\subsection{From Interstellar Dust to Comets}
A major advance in our understanding of comets in the 
20th century was made by the space probes Vega 1 and 2 
and Giotto (see Nature, comet Halley issue, vol. 321, 1986). 
Until that time no one had ever seen a comet nucleus. 
The critical new discoveries were: 
(1) the low albedo ($\approx 0.04$) of comets; 
(2) the size distribution of the comet dust extending down to 
interstellar dust sizes (10$^{-15}$ -- 10$^{-18}\g$), 
(3) the organic fraction of comet dust
(see Greenberg \& Li 1999b and references therein). 
The current ground based observations of the volatile 
composition of comets implies a close connection with 
the ices of interstellar dust (see Crovisier 1999 and
Irvine \& Bergin 2000 for recent reviews). 

Most of the current models of comet nuclei presume that 
to a major extent they are basically aggregates of the 
interstellar dust in its final evolved state in the collapsing 
molecular cloud which becomes the protosolar nebula. 
In addition to the chemical consequences of such a model, 
there is the prediction of a morphological structure in which
the aggregate material consists of tenth micron basic units 
each of which contains (on average) a silicate core, 
a layer of complex organic material, and an outer layer 
of ices in which are embedded all the very small carbonaceous 
particles characterizing the interstellar UV hump 
and the far UV extinction. All these components 
have been observed in the comet comae in one way 
or another.\footnote{%
  PAH molecules, a significant constituent of interstellar 
  dust (see \S3.2.2), would also be present in comets if 
  they indeed contain unprocessed interstellar matter.
  The presence of PAHs in comets has been suggested by 
  the 3.28$\um$ emission feature detected in some comets
  (Bockel\'{e}e-Morvan, Brooke, \& Crovisier 1995). 
  More specifically, a 3-ring PAH molecule 
  -- phenanthrene (C$_{14}$H$_{10}$) -- has been 
  proposed as the carrier for the 342--375$\nm$ fluorescence 
  bands seen in comet 1P/Halley (Moreels et al.\ 1994).
  Li \& Draine (2002d) are currently working on this topic. 
  }
The implication is that space probes which can examine in detail 
the composition of comet nuclei will be able to provide us with 
hands-on data on most of the components of interstellar dust 
and will tell us what is the end product of chemical evolution 
in a collapsing protosolar molecular cloud. At this time many
laboratories are preparing materials as a data base for 
comparison with what will be analyzed during the space missions.

\section{Future}
There are quite a few unsolved or partially solved problems 
related to interstellar dust which will be demanding close 
attention in the future (see below for a list). 
A number of new remote observational facilities 
which will be available early in the new millennium
(Atacama Large Millimeter Array [MMA/LSA],
 Far Infrared and Submillimeter Telescope [FIRST],
 Next Generation Space Telescope [NGST],
 Space Infrared Telescope Facility [SIRTF],
 Stratospheric Observations for Infrared Astronomy [SOFIA],
 Submillimeter Wave Astronomy Satellite [SWAS])
will permit further tests of current dust models and 
promise new observational breakthroughs.

\begin{enumerate}
\item What is the source and nature of 
      the Diffuse Interstellar Bands?

\item What is the carrier of the 2175$\Angstrom$ extinction hump?

\item What is the carrier of the ``Unidentified Infrared Bands''?
      if it is PAHs, where are they formed?
      are they mainly from carbon star outflows or formed
      in situ by ion-molecule reactions (Herbst 1991)
      or from the organic refractories derived from 
      photoprocessing of ice mixtures (Greenberg et al.\ 2000)?  

\item What is the carrier of the Extended Red Emission? 

\item What are all the sources and sinks (destruction) of 
      interstellar dust?
      where are interstellar grains made? are they mainly
      made in the cold ISM (Draine 1990) 
      or are the silicate cores mainly stardust (serving as 
      ``condensation seeds'') while the organic mantles 
      are formed in the ISM (Greenberg 1982a; 
      Greenberg \& Li 1999a)? 

\item What are the exact composition and morphology of 
      interstellar dust? are they separate bare silicate and
      graphite grains or silicate core-carbonaceous mantle
      grains or composite grains composed of small silicates, 
      carbon and vacuum? if most of interstellar grain 
      mass is condensed in the cold ISM, how can pure
      silicate and graphite grains form (see Draine 1995)?
        
\item What are the sizes of large dust grains
      ($> 0.25\um$)? how much can we learn from
      X-ray halos and from spacecraft in situ dust
      detections?

\item Why are crystalline silicates not seen in the ISM
      while they are present in stardust and cometary dust?
      how do cometary silicates become crystalized?

\item How do molecular hydrogen and other simple molecules
      form on grain surfaces?
      although considerable progress has been made 
      in recently years in studies of the diffusion rates 
      of adsorbed hydrogen atoms on the surfaces of variable 
      dust materials, the recombination reactions, 
      and the restoration of the new molecules to 
      the gas phase (Pirronello et al.\ 1997;
      Pirronello et al.\ 1999; Manic\`{o} et al.\ 2001),
      the formation of molecular hydrogen is still not
      well understood (Herbst 2000; Pirronello 2002). 

\item How do interstellar grains accrete and
      deplete mantles in dense molecular clouds? 
      we need high spatial resolution observations of
      molecule distributions in the gas and in the solid 
      as function of depth in the cloud -- interiors of
      clouds as well as regions of low and high mass star formation.

\item How does dust evolve in protosolar regions? 
      we need higher spatial resolution and sensitivity.
      Improvements in the theory of dust/grain chemistry, 
      particularly in collapsing clouds leading to
      star formation as well as in quiescent molecular clouds. 

\item Will the chemical and morphological analysis of comet nuclei 
      and dust material reveal the true character of interstellar 
      dust? will they provide further answers to the question of 
      life's origin?

\item How can we resolve the evolution of interstellar matter 
      leading to the material measured and analyzed in meteorites, 
      in interplanetary dust particles?

\item What is the true atomic composition of 
      the interstellar medium? 
      how variable is it in time and space?
      are there global variation over distances of kiloparsecs?

\item When did dust first form in a galaxy?
      what are the composition and sizes of dust 
      in extragalactic environments? 

\end{enumerate}

\centerline{\bf Acknowledgements}
\vspace{1mm}
\noindent{A. Li was deeply saddened by the passing away
of Prof. J. Mayo Greenberg on November 29, 2001. 
As a pioneer in the fields of cosmic dust, comets, 
astrochemistry, astrobiology and light scattering,
Mayo's passing was a great loss for the astro-community.
Mayo had been scientifically active till his very
last days. Just a few weeks before Mayo passed away,
A. Li discussed future collaboration plans 
with him on dust in high-$z$ galaxies.
It was a great experience for A. Li to 
work with Mayo in Leiden.
He will be remembered forever, 
as a great astrophysicist and as a great mentor.
A. Li is also grateful to Profs. Bruce T. Draine 
and Ewine F. van Dishoeck for their continuous advice, 
encouragement and support. 
A. Li thanks Profs. Lou J. Allamandola, Bruce T. Draine,
Jonathan Lunine, Valerio Pirronello for valuable 
discussions, comments and suggestions.
Some of the materials of \S2 (dust history) were taken 
from Greenberg \& Shen (1999) and the ``Introduction'' 
Chapter of A. Li's PhD thesis (Leiden, 1998);
of \S3.2.1 (dust luminescence) from Li \& Draine (2002a);
and of \S3.2.2 (dust IR emission) from Li \& Draine (2001b). 
This work was supported in part by NASA grant NAG5-7030 and 
NSF grant AST-9988126 and by a grant from the Netherlands 
Organization for Space Research (SRON).}

\end{document}